%% file: gn26.tex
\documentclass[11pt]{article}
\usepackage{amsmath,amsfonts,amssymb,fullpage,xcolor,graphicx,natbib,url,adjustbox,comment,pdfpages,soul}
\usepackage{subcaption}

\newtheorem{lemma}{Lemma}

\newtheorem{proposition}{Proposition}

\def\pconv{\smash{\mathop{\longrightarrow}\limits^{p}}}     %Converges in proby

\def\var{\text{var}}
\def\cov{\text{cov}}

\def\plim{\text{plim}}
\def\dconv{\smash{\mathop{\longrightarrow}\limits^d}}     
\def\pconv{\smash{\mathop{\longrightarrow}\limits^{p}}}     
\let\tilde\widetilde

\def\hi{w}
\def\stoc{\mu}

\begin{document}

\let\Large\large
\let\large\normalsize
%\title{Cross-Section Regressions in  Time Compressed Data: \newline Estimating the Economic Effects of Warming Temperature} 
\title{Cross-Section Estimation of Long-Run Relations Using Time-Compressed Data}

\author{ Serena Ng\footnote{Dept. of Economics, Columbia University and NBER.  1126D-420 W. 118 St., New York, NY 10025, Email: serena.ng at columbia.edu} \and Nikolay Gospodinov\thanks{Research Department, Federal Reserve Bank of Atlanta, Email: nikolay.gospodinov@atl.frb.org  \smallskip \newline Financial Support from the National Science Foundation (SES  2018369) is gratefully acknowledged. The views expressed here are the authors' and not necessarily those of the Federal Reserve Bank of Atlanta or the Federal Reserve System.} }
\date{\today}
%\\ Preliminary and Incomplete\\ Please do not Quote}
\maketitle
\bibliographystyle{harvard}
\begin{abstract}
%Many long-run relations such as between economic and demographic involve non-stationary variables, and inference from single equation or panel regressions  is often non-standard. 
Many empirical investigations of long-run relations   are  based on cross-section regressions in  averaged  or long differenced data that  effectively have the time dimension of a $T\times N$ panel compressed.  We analyze  a class of  time-compressed I(1) data and show that they have magnified variability stemming from the fact that   the cross-section variance   of a  non-stationary panel `fans out'  with time.   Cross-section regressions in time compressed data can potentially  yield   estimates that are super-consistent and asymptotically normal,  whether the regressors are stationary, non-stationary, or  highly persistent.  The fastest convergence rate of $\sqrt{N}T$ requires a compression scheme that not only magnifies the non-stationary signal,  but also dilutes the regression noise.  Omitted fixed effects preclude noise dilution but the estimates remain super-consistent.  However,  the fanning out effect can be weakened when the data have a strong force for mean-reversion or convergence, a problem that seems relevant for temperature data. We consider three applications and find that the long-run relation between consumption and income, and between growth/inflation and  demographic variables are reasonably well determined, but  the estimated relation between growth and  warming temperature is fragile.

\end{abstract}

\bigskip


\bigskip


\thispagestyle{empty}
\setcounter{page}{0}
\baselineskip=18.0pt

\newpage

\section{Introduction}
Many long-run economic relations are of interest:-    between consumption and  output,  productivity growth and real interest rates, public debt and GDP, purchasing power parity,   inflation and interest rate,  inflation  and demographics,   interest rates of different maturities. While it is common to study these relations  using  time series or panel data estimators with known theoretical properties,\footnote{See, for example,  \citet{pesaran-etal:07},  \citet{lunsford:17},  \citet{yi-zhang:17}, \citet{yoon-kim-lee:14}, and \citet{juselius-takats:21}.} cross-section regressions using  data averaged over long periods of time have also been employed. Examples include investigation of savings and investment  in \citet{feldstein-horioka},  growth and its determinants considered in \citet{feldstein-horioka, barro-91, easterly-levine} and \citet{mendelsohn-etal:94}. Long-differences of averaged data are also increasingly used  to estimate `longer run' relation between economic and environmental variables, as in \citet{dell-jones-olken:12, burke-emerick:16}. The data after averaging and differencing used in these regressions  essentially have no time dimension, or   what we will refer to as {\em time compressed}.    What are the properties of the time compressed data? Is one method better than another? How effective are such data  in recovering long-run relations?   Would  non-stationarity  matter,  and more generally, do time series properties of the data affect cross-section regressions?  This paper provides such an analysis.

Our point of departure is the result in 
\citet{deaton-paxson-jpe94} that income inequality will increase with age  if individual incomes are random walks,   with the implication
that the cross-section variance within a cohort will `fan-out' with age. Built on this insight, we study the properties of a class of time compression functions for I(1) (non-stationary) data and  obtain three results. First, the  cross-section variance of  data driven by simple stochastic trends will `fan-out' (or magnified)  over time.   The maximum fanning out  is   obtained by `point sampling' the last row of   a $T\times N$    panel, while  averaging the data over $K$ periods has a shrinkage factor  of one-third. Different   forms of non-stationarity may change the  fanning out profile,  but  significant  variation in  cross-section variance can be expected.

Second, the magnified signal from fanning out can be exploited in cross-section regressions. Lemma \ref{lemma:smooth} shows that  in the local-to-unity framework, the variance  of time compressed data traverses smoothly from $O(1)$ to $O(K)$ as  the largest autoregressive root moves  towards the unit circle.  The convergence rate of the least squares estimator self-adjusts with the degree of persistence.
Even though cross-section regressions do not explicitly use time variations, fast convergence is a direct consequence of the strong signal in  highly persistent data.   Third, we show that   super-consistent  (faster than $\sqrt{N}$) and asymptotically normal estimates  are usually possible. Proposition \ref{prop:prop1} shows that  the fastest convergence rate  requires that  time compression  performs both signal magnification and noise dilution.  While K-averaging achieves this rate, point sampling does not because of the inability to dilute noise.

The main appeal of cross-section regressions in time compressed  data is that  standard normal inference can be used without knowing  whether the regressors are I(1) or I(0),  with no need for HAC standard errors. In contrast, the asymptotic distributions from   time series or panel regressions  are either discontinuous if  the regressor has a unit root, and/or have nuisance parameters that complicate inference. The cross-section estimates are also  robust to  endogeneity and low-order misspecification biases because of  the magnified signal in the regressors.    Cross-section regressions based on time averaged data thus  have more desirable theoretical properties than previously understood.  

We use three applications to illustrate empirically relevant issues. In the first application that estimates the long-run relation between consumption and income using data for 44 countries from the Penn World Table,   omitted fixed effects prohibit noise dilution and preclude the fastest convergence rate, but super-consistent and asymptotically normal estimates can still be attained at the cost of an inflated variance. In the second application, we analyze the relation between  GDP growth or inflation taken from the Penn World Table,  and demographic variables taken from the World Bank database. The cross-section variance of the dependency ratio `fans-in'.  This feature  can be  traced to convergence across countries of the under-16 population,  in spite of clear fanning out of the share of population over 65. In such a case, it is informative to estimate the  effect of the two observed components separately. In the third application of growth on temperature using data for the 48  states in the U.S.,   the trending component is weak relative to the  stationary component, and the cross-section variance of temperature shows little fanning out. Unless the two latent components have the same effect on economic outcomes,   time compression schemes will be challenged  to disentangle the effect of   warming trend from regular temperature fluctuations. In this case, the    long- and short-run estimates  may well appear similar,  a finding that has been attributed to a lack of adaptation in the literature.  Checking the stability of the estimates with respect to $K$ is effective in gauging  whether the long-run relation of interest can be reliably estimated.

The paper proceeds as follows.  Section 2 begins with the key insight in \citet{deaton-paxson-jpe94} and Section 3 proceeds to analyze a class of time compression functions that include the important cases of  point-sampling and  $K$-averaging but exclude long-differencing. Section 4 considers estimation of a static model by cross-section regression in time compressed data under specific assumptions.  Section 5 relaxes  these assumptions. Section 6 considers mechanisms that can dampen the fanning out in the context of three applications.  Section 7 concludes.

\section{Fanning Out of Cross-Section Variations}

This section studies the cross-section 
 variability  in  a panel of $N$   non-stationary series observed over $T$ periods. We will  let $\mathbb E_i[\cdot]$ and $\var_i[\cdot]$  denote expectation and variance over the cross-section distribution of units, while $\var(\cdot)$ is variance for a given $i$.

Our point of departure is the result  in  \citet{deaton-paxson-jpe94}  that if  utility is quadratic and  the permanent income hypothesis (PIH) holds,  consumption  and income inequality will increase with age `$a$'.  The result arises because  consumption ($c$) is a random walk under PIH, and can be characterized as  $c_{ia}=c_{i,a-1}+u_{ia}$. Assuming in addition that  $\cov(c_{i,a-1},u_{ia})=0$,    the cross-section variance of  a given set of  $N$ individuals in the same cohort at age  $a$ is related to the one at age  $a-1$ by $\var_{i|a}(c_{ia})=\var_{i|a-1}(c_{i,a-1})+\sigma^2_i$.    It follows 
that  the cross-section variance of the  cohort  will increase with $a$, reflecting the cumulative difference in individual shocks as they aged.  \citet{ng:08} shows that if a fraction $\kappa$ of the panel is non-stationary, the cross-section variance will still fan out at rate $\kappa t$. 
\citet{deaton-paxson-jpe94} noted that other  preferences may increase or decrease consumption dispersion over time, and it  can even be nullified  if individuals are homogeneous. 
We now explore the fanning out effect outside of the PIH context, and under more general conditions.

\subsection{Generalized Fanning Out}
\begin{figure}[!ht]
\caption{log Real GDP Per-Capita: 44 Countries in PWT}
\label{fig:app05_gdp}
\hspace*{-.75in}
\centering
\includegraphics[width=8.0in,height=2.50in]{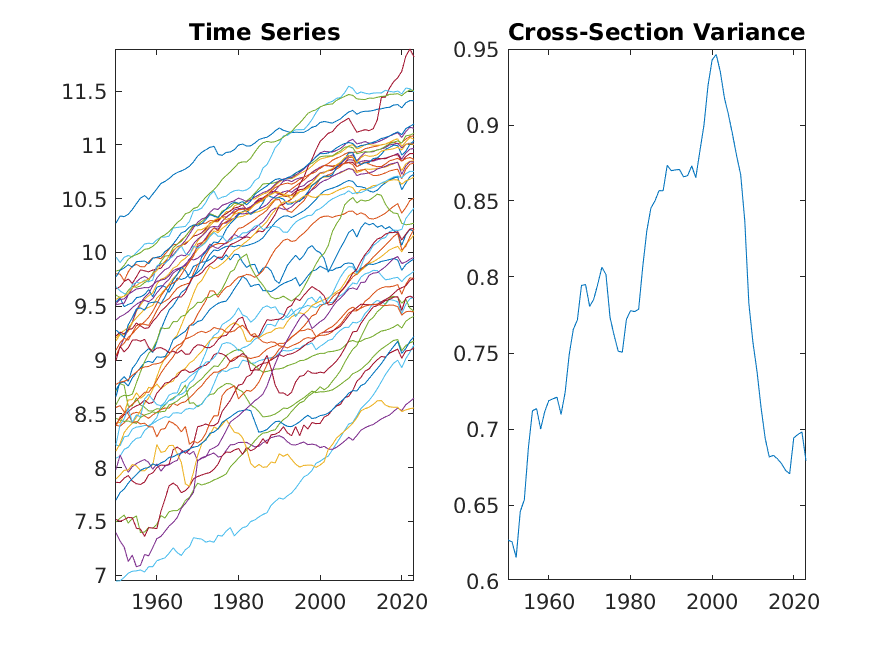}
\end{figure}

Our working assumption is that the data are non-stationary but  not necessarily a random walk. Consider
 log per-capita real GDP for 44 countries over the sample 1950-2023 taken from the Penn-World Table,  \citet{pwt110}. The  left panel of Figure \ref{fig:app05_gdp} shows that  the time series  are non-stationary,   and the right panel shows that the cross-section variance fans until 2000 when $\sigma$-convergence began.  Many national accounts variables in this database have this  property. Ratios such as  export-to-GDP, import-to-GDP, debt-to-GDP   in the Global Macroeconomic Database provided in \citet{gmd} also have  cross-section variances that fan out over a long span.

To study  the trending cross-section variance, we start with a simple data generating process.  For $t=1,\ldots  T$ with $x_{i0}=0$, let $\Delta x_{it}=u_{it}$, $u_{it}=C_i(L) \epsilon_{it}$, where  $\epsilon_{it}\sim (0,\sigma^2_i)$ is a white noise error with variance $\sigma^2_i$ that is independent of $C_{ij}$ for all $j$, $C_i(L)=1+C_{i1} L+C_{i2}L^2+\ldots$ is a polynomial in the lag operator $L$ satisfying  $\sum_{j=0}^\infty C_{ij}^2<\infty$.  Using $C_i(L)= C_i(1)+(1-L)C_i^*(L)$, where $C_i(1)=\sum_{j=0}^\infty C_{ij}$ and $C_{ij}^*=-\sum_{k=j+1}^\infty C_{ik}$,  the Beveridge-Nelson Decomposition  yields  
\begin{equation*}
x_{it}=C_i(1)\sum_{s=1}^{t}\epsilon_{is}+C_i^{\ast }(L)\epsilon_{it}=C_i(1)\sum_{s=1}^{t}\epsilon_{is}+w_{it}.
\end{equation*}
In the above,  
 $x_{it}$ is decomposed into a random walk (non-stationary or I(1)) component $C_i(1)\sum_{j=1}^{t}\epsilon_{ij}$, and a mean-reverting (stationary or I(0)) component $w_{it}=C_i^*(L)u_{it}$.
For a given $i$, 
\[
\var(x_{it})= C_i(1)^2 \sigma^2_{i} t + \var (w_{t}),
\]
where $C_i(1)^2\sigma^2_i=\omega^2_i$ is the long-run variance of $\Delta x_{it}$. Thus, unlike an I(0)  process when $C_i(1)=0$ and $\var(x_{it})=O(1)$, the variance of an I(1) process  increases with calendar time $t$. It follows that   the cross-section variance   fans-out at rate $t$:
\[ \var_{i}(x_{it})= \mathbb E_{i}[\omega^2_i]\cdot t + \var_i[w_{i,t}]=O(t).\]
Fanning out  requires   $\mathbb E_i [\omega^2_i]\ne 0$, not that every $\omega^2_i>0$.

  Introducing serial correlation in $u_{it}$ changes the pre-factor of the fanning out rate from the average short-run variance $\mathbb E_i[\sigma_i^2]$ to the average long-run variance $\mathbb E_i[\omega^2_i]$. This, however, is not inconsequential. Consider  the  IMA(1) model $\Delta x_{it}=\epsilon_{it}+\vartheta_i \epsilon_{it-1}$. With a long-run variance of $\omega^2_i=(1+\vartheta_i)^2\sigma_i^2$,  the cross-section variance at $t$  is 
\[\var_{i}( x_{it})=\mathbb E_i[(1+\vartheta_i)^2\sigma^2_i] \cdot t+\var_i[\vartheta_i^2\sigma^2_i].\]
 The closer is $\mathbb E_i[\vartheta_i]$ to -1, the weaker is the fanning out. Two models with IMA(1) as reduced form are: i). the factor model  $x_{it}=\lambda_i f_t+v_{it}$, where $f_t=f_{t-1}+u_t$ with  $\var_i(x_{it})=\var_i(\lambda_i)f_t^2+\var_i(v_{it})$; and ii). the local level model with $x_{it}=\mu_{it}+v_{it}$, $\mu_{it}=\mu_{it-1}+u_{it}$.  In both models,  $v_{it}$ is stationary.

\subsection{Convergence and Initial Conditions}
The above simple non-stationary structure may not be a good characterization of some  socio-economic data.  For example, the cross-section variance in Figure \ref{fig:app05_gdp} stops  increasing around year 2000 when emerging and populous economies have grown much faster than matured ones.   We now consider alternative forms of non-stationarity and show that  some features can interact to generate complex  cross-sectional variability.

\begin{itemize}
\item[i] There may be situations when it is  appropriate to assume $x_{i,-T_{int}}=0$ instead of $x_{i0}=0$.  For example, if $x_{i1}$ is wealth at the start of retirement,  then $x_{i,-T_{init}}=0$  when unit $i$  entered the labor force  $T_{\text{init}}$ periods earlier. The effective cross-section variance at  $t$ is  $O(\mathbf t)$, where $\mathbf t=(t+T_{init})$.  
% If $x_{i,-T_{init}}=0$, then with  $\mathbb T_1=T_{init}+T_1$,  point sampling gives $\var_i(x_{i,T_1})=\mathbb T_1 \mathbb E_i[\sigma^2_i]$.
\item[ii] Suppose $\Delta x_{it}=g_i+u_{it}$, where $g_i$ is an exogenous drift. From $x_{it}=x_{i0}+g_t t+\sum_{s=1}^t u_{is}$,  
\[\var_i(x_{it})=\var_i(x_{i0})+ \mathbb E_i[\omega^2_i] t+ \var_i(g_i) t^2=O(t^2).\]
The faster fanning out rate  of $t^2$ arises because the  deterministic trend dominates  the stochastic trend. 
Nonetheless,  heterogeneity in $g_i$ is needed to achieve  the faster rate. 
\item[iii] Suppose that $x_{i0}\ne 0$ and   $\cov_i(x_{i0},g_i)< 0$. Then,
  \begin{eqnarray*}
  \var_i(x_{it})&=&\var_i(x_{i0})+ \bigg(2  \cov_i(x_{i0},g_i)+  \mathbb E_i[\omega^2_i]\bigg) t+ \var_i(g_i) t^2\\
  &=& \var_i(x_{i0})+c_1 t + c_2 t^2=O(t^2).
  \end{eqnarray*}
If $\mathbb E_i[\omega^2_i]< -2\cov_i(x_{i0},g_i)$, then  $c_1<0$ and  $\var_i(x_{it})$ may trend down for some values of $t$. But this negative trend  can only be temporary because $\var_i(g_i)t^2$ grows with $t$. In such cases, the cross-section variance could exhibit a U-shape with a minimum at $t^*=-\frac{c_1}{2c_2}$. 

\item[iv]  Suppose $x_{it}=\lambda_if_t+ v_{it}$, $v_{it}$ is stationary,  and
 $f_t=g_f+ f_{t-1}+u_t$, where $g_f$ is the drift of the common factor $f_t$, and  $u_t\sim (0,\sigma^2)$ is stationary and has long-run variance $\omega^2$.   Thus,
  \[\var_i(x_{it})=\var_i(x_{i0})+\var_i(\lambda_i) f_t^2+\var_i(v_{it})=O(t^2).\]
But  $f_t=f_0+g_f t+\sum_{s=1}^t u_s$.  If $g_f<0$ and $f_0\ne 0$,    $f_t^2=(f_0+g_f t)^2+O_p(t)$ which is quadratic in $t$  
will be U-shaped, reaching a minimum at $t^*=-f_0/g_f$ before increasing. 
Even with heterogeneity in $\lambda_i$, the cross-section variance may not always increase with $t$. 
  %The Appendix shows that the cross-section variance of K-averaged data can also be U-shaped, but with different turning points.  

\end{itemize}

In general, the fanning out profile will depend on data and the sample.  There can be episodes when the cross-section variance will not increase, and could even be  $O(1)$.  To obtain general results, we will  assume that  $\Delta x_{it}=u_{it}$ where $x_{i0}$ is exogenous, with the understanding that the fanning out rate may be faster or slower than $t$ obtained under alternative assumptions.
%The bottom panel of Figure \ref{fig:fig0} plots demeaned temperature data for the 50 states in the U.S. The upward trend since the 1960s is notable, and yet the cross-section variance does not appear to be fanning out, suggesting that $\mathbb E_i[\omega^2_i]$ is small. 

\section{A Class of Time Compression Functions}
This section shows that the fanning out effect is a feature of a class of time-compressed data.  For each $i=1,\ldots T$ and $t=1,\ldots T$, let   $\Delta x_{it}=u_{it}$  with $x_{i0}=0$, and let
\[\tilde  X_{iT_1}=\sum_{t=T_0+1}^{T_1}  w(t) x_{it} .\] 
be defined by a time-compressed function $w(t)$  satisfying the following.  

\paragraph{Assumption A:} For $T_0\ge 0$ and $T_1\le T$,  (i) $w(t)\ge 0$ for all $t\in[T_0+1,T_1]$; (ii) $\sum_{t=T_0+1}^{T_1} w(t)=1$, and (iii) $w(t)>0$ for some $t\in[T_0+1,T_1]$.

Assumption A.(i) and A.(ii)  ensure that $\{w(t)\}$ is a proper probability distribution over the interval $[T_0+1,T_1]$.   A.(iii) ensures that the weights are not concentrated in the early sample.

 The class of functions   include {\em concentrated}  weights. An example  is   point sampling, a one parameter time compression scheme that takes   row $T_1$ from  a $T\times N$ panel:   
\[\tilde X_{iT_1}=x_{i,T_1}=x_{i0}+\sum_{j=1}^{T_1} u_{ij}=O(T_1).\] 
Point sampling thus has  $w(t)=1_{t=T_1}$, and puts a  weight of  $\phi_j=1$ on  innovation $u_j$ for all $j\le T_1$.
Assumption A.(iii) requires that $T_1$ is not too close to the origin so that enough innovations are accumulated to generate fanning out.
The preceding analysis implies that  variability of  $\tilde X_{iT_1}$ is maximized by point sampling at $T_1=T$.   It is also immediate that  $\var(\tilde X_{iT_1})= O(1)$ if $x$ is stationary, so the cross-section variance neither fans in nor out.

Whereas fanning out in point-sampling means that $\var_i(x_{it})$ increases with calendar time $t$, as in \citet{deaton-paxson-jpe94},  fanning-out using other schemes will be taken to mean that $\var_i(\tilde X_{iK})=O(K)$, where $K=T_1-T_0$ depends on at least two  choice parameters, $T_0$ and $T_1$.
Such functions include  {\em spread-out} weights, 
 the simplest being full-sample averaging with $T_0=0$, $T_1=T$. But it also includes  a three parameter scheme (referred to in the simulations below as DA) that  gives a weight of zero to the first  $\kappa$ fraction of the sample, and  the remaining $1-\kappa$ fraction  would receive a weight of $\frac{1}{(1-\kappa) K}$, with $K=T_1-T_0$.

\subsection{A Closer Look at $K$ Averaging}
Many well-cited cross-section regressions are based on time averaged data. For example, \citet{feldstein-horioka} averaged the data over 15 years, while \citet{barro-91} averaged the data over 25 years.   The WMO standard for long-run temperature is 30 years.\footnote{See \url{https://wmo.int/media/news/updated-30-year-reference-period-reflects-changing-climate}.}  It is thus useful to take a closer look at averaging data over $K$ periods.

We will use $\bar X_{iK}$ to denote time-compressed data by averaging. Averaging observations over time   is  commonly  thought to remove high frequency  noise,  making it easier to visualize patterns over time.  Consider first full-sample averaging where  $\bar X_{i,T}=\frac{1}{T}\sum_{t=1}^T x_{it}$. If $x_{it}$ is stationary  with long-run variance $\omega^2_i$,  then by central limit theorem, $\sqrt{T}\bar X_{i,T}\dconv N(0,\omega_{i}^2)$. Time averaging of stationary data reduces noise  relative to  an arbitrary draw $x_{it}$ at any $t$. It follows from $\var(\bar X_{i,T})=O(T^{-1})$ that the cross-section variance $\var_i(\bar X_{iT})=O(T^{-1})$ `fans-in'.

Consider now the I(1) case when  $\Delta x_{it}=u_{it}$, where $x_{i0}=0$,  $u_{it}\sim (0,\sigma^2_{i})$ is stationary with long-run variance $\omega^2_i$. Then by functional central limit theorem,
$  T^{-1/2} \bar X_{i,T}\Rightarrow \omega^2_i\int_0^1 W(r)dr$, where $W(r)\sim N(0,r)$ is a standard Brownian motion. Since
$E[\int_0^1 W(r)dr]^2=1/3$,  for large $T$, we have
$\var(\bar X_{i,T})=\omega^2_i\frac{T}{3}. $
In contrast to the stationary case when  $\var_i(\bar X_{iT})$ fans-in,   now
\[\var_i(\bar X_{iT})=\mathbb E_i[\omega^2_i]\frac{T}{3}\] 
 fans-out, but shrinks by 1/3 relative to  point sampling at $T$.

To understand the reason for this shrinkage,  consider  averaging $K=T_1-T_0$ observations between $T_0+1$ and $T_1$, which may be needed if complete data is not available for the full sample, or properties in the sub-sample are of interest. When applied sequentially to a single time series, $K$-averaging  is a low-pass filter that  passes all variations with  periodicities longer than $K$ and attenuates shorter ones.\footnote{ Let $\omega=2\pi/K$. The discrete Fourier transform of the  response is $H(w)=\frac{1}{K}\sum_{m=0}^{K-1} e^{-j\omega m}=\frac{1}{K} \frac{1-e^{-j\omega K}}{1-e^{-j\omega}}$. Using $\sin(\theta) = \frac{e^{j\theta}-e^{-j\theta}}{2j}$, the frequency response is the Dirichlet kernel, $H(\omega)\approx \frac{\sin(K\omega/2}{K\omega/2}$ whose  first zero is at $\omega=2\pi/K$.}  We use $K$-averaging  only once for each $i$   to generate
\[\bar X_{iK}= \frac{1}{K}\sum_{t=T_0+1}^{T_1} x_{it}=x_{i,T_0}+ \frac{1}{K}\sum_{t=T_0+1}^{T_1}\sum_{s=T_0+1}^{t}  u_{is}.\]
Simplifying  the double sum, we   obtain 
\begin{eqnarray*} 
\bar X_{iK}&=&x_{i,T_0}+\sum_{j=1}^{K}\bigg(\frac{K-j+1}{K}\bigg) u_{i,T_0+j}=x_{i,T_0}+\sum_{j=1}^{K} \phi_j u_{i,T_0+j},
\end{eqnarray*}
where $\phi_j=\frac{K-j+1}{K}$ is the weight on innovation $u_{i,T_0+j}$, and     $\sum_{j=1}^K \phi_j^2 =\frac{K(K+1)(2K+1)}{6K^2}$. Note that  the relevant initial condition for $K$ averaging is $x_{i,T_0}$, not $x_{i0}=0$. 

%\footnote{The law of total variance holds that $ \var_i(\bar X_{iK})=\mathbb E_i[\var(\bar X_{iK}|\sigma^2_i)] +\var_i(\mathbb E[\bar X_{iK}|\sigma^2_i]$, and $\mathbb E[\bar X_{iK}]=x_{i,T_0}$}

\begin{lemma}
\label{lem:lemma1}
Suppose that $x_{it}=x_{it-1}+u_{it}$ where $u_{it}\sim (0,\sigma^2_{i})$   is independent across $i$ but possibly serially correlated with long-run variance $\omega^2_i$,    and $x_{i0}=0$. Let $\bar X_{i,K}$ be the mean of unit $i$ computed from  $K$ observations between $T_0$ and $ T_1$.   Let $ \bar a_K=\frac{(K+1)(2K+1)}{6K}$. For fixed $K$,
\[ \var(\bar X_{iK})=\var_i(x_{i,T_0})+\bar a_K\omega^2_i,\]
and for large $K$ relative to $T_0$,
 $\var_i(\bar X_{iK}) =  \mathbb E_i[\omega_i^2] K/3$. \hspace{.25in} $\square$
\end{lemma}

The cross-section variance has two components: one from the variance of the conditional mean, which in this case is the initial condition,   and one from the mean of the conditional variance due to the shocks. Now $\var_i(x_{i,T_0})=\mathbb E_i[\omega_i^2] T_0$, and assuming $T_0$ is small relative to $K$ gives  $\var_i(\bar X_{iK})\approx \mathbb E_i[\omega^2_i]K/3$ since $\bar a_K/K\approx 1/3$ for large $K$.

 \subsection{Properties of $\tilde X_{iK}$}
The results for point sampling and $K$-averaging draw attention to two distinct sets of weights of interest: $\{w(t)\}$ on the data, and $\{\phi_s\}$  on the innovations. This will still be the case as
we now analyze the properties of $\tilde X_{iK}$ for any $w(t)$ satisfying Assumption A.

In the I(1) case, $x_{it}=x_{it-1}+u_{it}= x_{it}=x_{i0}+\sum_{s=T_0+1}^t u_{is}$.  Changing the order of summation to sum over $s$ first and then for $t>s$, we obtain
\begin{eqnarray*}
\tilde X_{iK}&=& x_{i,T_0}+\bigg(\sum_{t=T_0+1}^{T_1} w(t)\sum_{s=T_0+1}^t u_{is} \bigg)=x_{i,T_0}+\sum_{s=T_0+1}^{T_1} u_{is}\sum_{t=s}^{T_1} \ w(t)\\ &=&x_{i,T_0}+\sum_{s=T_0+1}^{T_1}  \phi_s  u_{is}, \quad \quad \text{with} \quad\quad \phi_s=\sum_{t=s}^{T_1} w(t).
\end{eqnarray*}
  Note that   $ \phi_s$ is the tail sum of $w$ at  $s$, and the sequence  $\{ \phi_s\}$ is  the survival function of the weight distribution $ w$,  
  starting at $ \phi_{T_0+1}=\sum_{t=T_0+1}^{T_1}w(t)=1$ and decreases  to $\phi_{T_1}= w(T_1)\ge 0$. Then,
\[\var(\tilde X_{iK})=\var_i(x_{i,T_0})+\sigma^2_i\sum_{s=T_0+1}^{T_1}\phi_s^2\equiv \sigma^2_i \Phi_K, \quad\quad \Phi_K=\tilde a_K(w) K.\] 
Condition A.(iii) ensures that $\sum_{s=T_0+1}^{T_1}\phi_s^2 >0$ for any $K$. In general, $\Phi_K=O(K)$ but the precise magnitude of fanning out depends on  $\tilde a_K(w)$ which is scheme specific. Full sample averaging has $\tilde a_K=1/3$, while  point sampling  has $\tilde a_K=1$.

In the I(0) case,  we have, by definition,
\begin{eqnarray*}
\var(\tilde X_{iK})&=& \var_i(x_{i,T_0})+\sum_{s=T_0+1}^{T_1}\sum_{t=T_0+1}^{T_1} w(t) w(s) \cov(x_{it},x_{is}).
\end{eqnarray*}
 %From the properties of covariance stationary processes,
 %it is immediate  that with $K$ period averaging, $\var(\tilde X_{iK})=O(K^{-1})$. To gain more insight,  
To make progress, consider  the AR(1) case where $x_{it}=\rho x_{i,t-1}+u_{it}=\rho_i^t u_{i0}+\sum_{s=0}^\infty \rho_i^s u_{t-s}$  and $u_{it}\sim (0,\sigma^2_i)$. With  $\cov(x_{it},x_{is})=\frac{\sigma^2_i}{1-\rho_i^2} \rho_i^{|t-s|}$ and assuming $\rho_i^t$ is small for large $t$, we have
\[\tilde X_{iK}=\sum_{s=T_0+1}^{T_1} \psi_{s}(\rho_i) u_{is},  \quad\quad \psi_s(\rho_i)=\sum_{t=s}^{T_1} w(t) \rho_i^{t-s}.\]
 The function $\psi_s(\rho_i)$ plays the role of $\phi_s$  in the unit root case.  We show in the Appendix that
\begin{eqnarray*}
 \var(\tilde X_{iK})&=&\frac{\sigma^2_i}{1-\rho_i^2} \sum_{h=-(K-1)}^{(K-1)} \sum_{s,t, t-s=h} w(t) w(s) \rho_i^{|t-s|}\\
 &=& \frac{\sigma^2_i}{1-\rho_i^2} \sum_{h=-(K-1)}^{K-1} A_h \rho_i^{|h|}\equiv \frac{\sigma^2_i}{1-\rho_i^2} G_w \equiv \sigma^2_i \Psi_K(\rho_i),
\end{eqnarray*}
where $A_h=\sum_{s,t,t-s=h}^{T_1} w(t) w(s)$ is  the total weight placed on observations $s$ and $t$, separated $h$ periods apart.   
Under Assumption A,  $\sum_{h=-(K-1)}^{K-1} A_h =(\sum_{t=T_0+1}^{T_1} w_t)^2=1$, implying that  $\{A_h\}$ forms a probability distribution over periods separated by $h$.   Thus,
$G_w(\rho_i)=\mathbb E_{A}|\rho_i^{|h|} \in [0,1]$  is bounded for any  $\rho_i\in[0,1]$ and any $w(t)$. 
But  $\var(\tilde X_{iK})$ increases with $K$ when $x$ is I(1). To transition from a fixed $\rho_i<1$ to $\rho_i=1$,  it must be that  $\Psi_K=\frac{G_w}{1-\rho_i^2}$ `blows-up' with $K$  via $\frac{1}{1-\rho_i^2}=\frac{1}{(1-\rho_i)(1+\rho_i)}$ as $\rho_i$ approaches the unit circle.  In  the local-to-unity case when $\rho_i=1-\frac{c_i}{K}$,  $G_w=\mathbb E_A[\rho_i^{|h|}]\approx 1-\frac{c_i}{K}\mathbb E_A[|h|]=O(1)$ for any choice of $w(t)$ satisfying Assumption A and any finite $h$. It is
$ \frac{1}{1-\rho_i} \propto \frac{K}{c_i}=O(K) $ that
 allows $\Psi_K$ to diverge with $K$ as $\rho_i$ moves towards the unit circle.

\begin{lemma}
\label{lemma:smooth}
 Let $\tilde X_{iK}=\sum_{t=T_0+1}^{T_1}  w(t) x_{it}$,   where $w(t)$ satisfies Assumption A. For given $K$, it holds that
\begin{itemize}
\item[a.] $\psi_s(\rho_i) \;\xrightarrow{\rho_i\to 0}\; w(s)$ 
 and $ \psi_s(\rho_i) \;\xrightarrow{\rho_i\to 1}\; \phi_s$;
%\item[b.] $\Psi_K(\rho_i)\rightarrow \sum_t w(t)^2=O(K^{-d})$ as $\rho_i\rightarrow 0$ where $d=0$ for point sampling, and $d=1$ for averaging based schemes.
\item[b.]  $\Psi_K(\rho_i)\rightarrow \Phi_K=O(K)$ as $\rho\rightarrow 1$.
\end{itemize}
\end{lemma}
Part (a)  says that the tail sum $\phi_s$ interpolates continuously
between the two extremes. 
%Part (b) implies that for stationary processes with $\rho_i$ far from the unit circle, there is no transition as $\Psi(K)$ is $O(K^{-1})$ for averaging schemes and $O(1)$ for point sampling. 
Part (b) indicates that $\Psi_K$  moves towards the unbounded $\Phi_K$ as $\rho_i$ moves towards the unit circle.   Indeed, if  $\rho_i=1-\frac{c_i}{K^d}$, then $\frac{1}{1-\rho_i}\propto \frac{K^d}{c_i}$ for any $d>1$. 
 This  allows $\var(\bar X_i)$ to transition smoothly from being bounded to divergent as persistence increases. This result is important to preserve continuity of the asymptotic distribution in regressions that use $\tilde X_{iK}$ as  regressor.

\subsection{Long Differencing}
We use $\breve X_{iK}$ to denote data transformed by long differencing (LD) constructed as in \citet{burke-hsiang-miguel:15}, \citet{dell-jones-olken:12}. Precisely, LD is based on the  difference of  two local averages of $M=2m+1$ observations, one  centered at $T_a$, and  one at  $T_b$: 
\[ \breve X_{iK}=\bar X_{iM}(T_b)-\bar X_{iM}(T_a).\]
 It thus requires three parameters, $m$, $T_a$ and $T_b$. Though quite widely used, the properties of LD  data are not well understood.\footnote{\citet{burke-emerick:16} simulate $x_{it}$ as having a unit-specific deterministic trend and  show that if $K$ is sufficiently large,  the LD estimates from a cross-section regression are more precise than the ones obtained from a $T\times N$ panel of raw data. 
%\citet{burke-emerick:16} simulated  $x_{it}=\lo t + \hi_{it}$ and $y_{it}=\alpha_{is}+ \beta_{\bar X} \bar X_i+\beta_\hi \hi_{it}+\varepsilon_{it}$. Their regression also has a quadratic term  that is not important for the arguments to follow.
}   If $m=0$, the LD(0,$T_a,T_b$) becomes a simple long difference,  $\breve X_{iK}= x_{i,T_b}-x_{i,T_a}=\sum_{s=T_a}^{T_b} u_{is}$.  Furthermore, if $x_{i0}=0$,  LD($0,1,T_b)$  is the same as a  point sampling at $T_1=T_b$  and is a member of our time-compression class. With a unit weight on all innovations between 1 and $T$, $\var_i(\tilde X_{iK})=O(K)$. \citet{cochrane-jpe:88} used the variance of a long-difference relative to a short difference to assess the relative importance of the common component.  For other parameterizations,   the LD weights on the innovations are not so transparent.  An example  helps.  Suppose   $T_a=4, T_b=10$, and $m=1$. The  observation and innovation weights  are:
\begin{center}
\begin{tabular}{lrrrrrrrrrrrr}
\hline
 &  & 3 & $4$ & $5$ & $6$ & $7$ & $8$ & $9$ & $10$ & $11$ & $12$\\
\hline
$\breve w(t)$  & $\cdot\cdot$ & $-\frac{1}{3}$ & $-\frac{1}{3}$ & $-\frac{1}{3}$ & $0$ & 0 & 0 & $\frac{1}{3}$ & $\frac{1}{3}$ & $\frac{1}{3}$ & 0 & $\cdot\cdot$\\
$\breve\phi_s$ & $\cdot\cdot $ & 0 &  $\frac{1}{3}$ & $\frac{2}{3}$ & $1$ & $1$ & $1$ & $1$ & $\frac{2}{3}$ &  
$\frac{1}{3}$ & 0 &$\cdot\cdot$  \\
\hline
\end{tabular}
\end{center}
We see that LD fails Assumption A because $\breve w(t)$ is not always positive and $\sum_t \breve w(t)=0$, not  1. Furthermore,  the innovation weights  $\{\breve \phi_s\}$ are not smooth; it  jumps from 0 to $1/M$ at $T_a-m+1$, and  from $1/M$ to 0 at $T_b+m+1$.   Perhaps the most notable feature of  LD is that it puts a weight of 1 on the
 $K-M+1$ innovations that are in the middle, while the observations at the two ends have smaller weights. This `tent' pattern of $\tilde\phi_s$ generalizes, and it can be shown that for $m>0$, 
%\footnote{We have used $2\sum_{j=1}^{2m} (j/M)^2= \frac{(M-1)(2M-1)}{3M}$ and $M=2m+1$ to simplify.}  %
\begin{equation}
\label{eq:eqLDV} \var(\breve  X_{iM})= \sigma^2_i\bigg((K-M+1) + 2\sum_{j=1}^{2m} (j/M)^2\bigg)= \sigma^2_i\bigg(K-\frac{(M^2-1)}{3M}\bigg)\approx \sigma^2_i(K-M/3).
\end{equation}
 Local averaging of $M$ observations  does not change the fact that the fanning out rate is determined by cross-section heterogeneity over $K$ periods. In this regard, LD isolates the same variations as point sampling $K$-averaging. However, when $M$ is large, the effective rate of $K-M/3$ can be much slower than $K$. 

The LD combines local-averaging (a low pass filter with zero at $\omega=\frac{2\pi}{M}$, known to attenuate high frequency variations) with differencing (a high-pass filter with zero at $\omega=0$, known to attenuate low frequency variations).\footnote{The squared gain 
$ H_{LD}(\omega)|^2\propto \bigg( \frac{\sin (M\omega/2)}{M\sin(\omega/2)}\bigg)^2 4\sin^2 (\frac{\omega K}{2})$ is a product of a function increasing from 0 at $\omega=2\pi/K$ and another function that decreases to zero at $\omega=2\pi/M$. The peak, $\omega^*$, thus lies between the two zones.}    It has a peak periodicity of  roughly   $p^*=\frac{2\pi}{\omega^*}=\frac{2MK}{M+K}$, which is the mean of the two boundary frequencies of $\frac{2\pi}{K}$ and $\frac{2\pi}{M}$. The choice of both $M$ and $K$ are important.
 However, the use of 'medium' and 'longer' terminology in the literature is loose.  \citet{dell-jones-olken:12} report `medium-term' effects using    $m=7$ and $K=15$ which implies $p^*=15$.  In contrast, \citet{burke-emerick:16} uses $m=2$ and $K=20$, which implies a lower periodicity of $p^*=8$ in spite of the `longer-term' terminology.

%It is useful to compare LD with $K$ averaging when both use the maximum span of  $T=K$. With $K$ period averaging,  this requires setting $T_0=1$ and $T_1=T$ to yield $\var(\bar X_{iK})=\sigma^2_i c_T$.  With LD, the largest  span is $T-2m$, obtained by  setting $T_0=m$ and $T_1=T-m$ to yield $\var(\breve X_{iK})=\sigma^2_i(T-M+1-\frac{M^2-1}{3M})$. The  LD variance is smaller than  $K$ averaging if and only if $M<3T/4$. If $M$ is larger,  there will not be enough observations that receive unit weighting to realize the variance advantage. The choice of $m$ is thus important.  

% We first consider models that specify $x_{it}$ as the sum of a stationary and a non-stationary component. We then specify a bivariate model where $x_{it}$ may not be stationary or exogenous.

\subsection{Numerical Examples}
We illustrate the extent of signal magnification by simulating data from  $x_{it}=\mu_{it}+v_{it}+\theta v_{it-1}$ where $v_{it}$ is stationary. Two models are considered: 
 \begin{itemize}
\item[1.]  local level model, where $\mu_{it}=\mu_{i,t-1}+u_{it}$ with $u_{it}\sim(0,\sigma^2_{u,i})$ and
 $q_i=\frac{\sigma^2_{u_i}}{\sigma^2_v}$; 
\item[2] factor model: $\mu_{it}=\lambda_{i} f_{t}$, where $f_t=f_{t-1}+u_t$, $u_t\sim(0,\sigma^2_u)$, $\lambda_{i}\sim (0,\sigma^2_\lambda)$  with $q=\frac{\sigma^2_\lambda \sigma^2_u}{\sigma^2_v}$.  
\end{itemize}
In both models,  shocks to $\mu_{it}$ are uncorrelated with $v_{it}\sim(0,\sigma^2_v)$. The local-level model provides a permanent-transitory decomposition of the data.   Since  $q_i=\frac{\sigma^2_{u_i}}{\sigma^2_v}$,  signal magnification requires  the average variability of $\mu_{it}$ to dominate those of the mean-reverting component $w_{it}$.
 The factor model  provides a common-idiosyncratic representation of the data. The   non-stationarity   common trend $\mu_t$  is not enough for signal magnification; it also requires    the loadings to be sufficiently heterogeneous.
Both have an IMA(1) model $\Delta x_{it}=u_{it}+\vartheta_i u_{it-1}$ as reduced form, where $ \vartheta_i=[-(q_i+2)\pm\sqrt{q_i^2+4q_i}]/2$ is closer to -1  the smaller the signal-noise ratio $q$ is.  
 \begin{table}[!ht]
\caption{Cross-Section Variance, $\var_i(\tilde X_{i,K})$: $(T_0,T)=(0,50)$}
\label{tbl:csvar}

\begin{center}

Local-Level Model: 

\begin{adjustbox}{width=\textwidth}

\begin{tabular}{ll|rrr|rrr|rrr|rrrr} \hline
 & & \multicolumn{3}{c}{PS:$\var_i(x_{i,T_1})$} &
   \multicolumn{3}{c}{KA($T_0,T_1)$} &
   \multicolumn{3}{c}{DA($T_0,T_1$)}  &
   \multicolumn{3}{c}{LD($m,T_a,T_b$)}  \\ \hline
$\sigma_u$ & $\vartheta$& $T_1=10$ & $T_1=30$ & $T_1=T$ & $T_1=10$ & $T_1=30$ & $T_1=T$ & $T_1=10$ & $T_1=30$ & $T_1=T$ & $(0,15,40)$ & $(2,15,40)$ & $(7,15,40)$  \\ \hline
\input table1_locallevel.tex
\hline
\end{tabular}
\end{adjustbox}

\bigskip

Factor  Model
\begin{adjustbox}{width=\textwidth}

\begin{tabular}{ll|rrr|rrr|rrr|rrrr} \hline

% & & \multicolumn{3}{c}{$\var_i(x_{i,t})$} &
%   \multicolumn{3}{c}{$\var_i(\bar X_{i,K})$} &
%   \multicolumn{3}{c}{$\var_i(\widetilde X_{i,K})$}  \\ \hline
$\sigma_\lambda$& $\vartheta$ & $T_1=10$ & $T_1=30$ & $T_1=T$ & $T_1=10$ & $T_1=30$ & $T_1=T$ & $T_1=10$ & $T_1=30$ & $T_1=T$ & $(0,15,40)$ & $(2,15,40)$ & $(7,15,40)$\\ \hline
\input table1_factormodel.tex
\hline

\end{tabular}
\end{adjustbox}
\end{center}
{\footnotesize Note: (i) PS is point sampling; (ii)  KA is K-averaging with  $w(t)=1/K$ for $t\in[T_0+1,T_1]$; (iii)  DA   averaging  uses $w(t)=0$ for $t\in[T_0+1,T_1-m]$ and $w(t)=1/m$ for $t\in[T_1-m+1,T_1]$, $m=K/2$; (iv) LD($m,T_a,T_b)$ is long-differencing of two $m$-period local averages centered at $(T_a,T_b)$. Throughout, $T_0=0$.}
\end{table}

Table \ref{tbl:csvar} reports the cross-section variance of four time compression schemes: (i) point sampling (PS), (ii) $K$-averaging (KA),  (iii) an averaging  scheme (DA) with $w_t=0$ for $t\in[T_0+1,T_1-m]$ and $w_t = 1/m$ for $t\in[T_1-m+1,T_1]$ where $m=K/2$, $K=T_1-T_0$, and (iv) long-differencing (LD). We evaluate LD$(m, T_a,T_b)$  at $T_a=m$ and $T_b=T_1$ to accommodate  the $m$  observations for the centered-moving average so that it is  comparable to PS at $T_1$.  The results  in the first panel confirms  that point sampling has more fanning out compared to KA and DA, the two compression schemes satisfying Assumption A. The second panel for $K$ sampling is in line with Lemma \ref{lem:lemma1} that the fanning out is about 1/3 of point sampling.  The third panel shows that by reducing the weights on the early observations, the fanning out effect of DA is larger than $K$-averaging that gives equal weight to all observations. The fourth panel shows that fanning out of LD is larger tha K-averaging, but  is smaller the larger is $M$.   The results also show that the signal magnification effect is smaller the closer $\vartheta$ is to -1.

\section{ Cross-Section Regressions in Time Compressed Data}

This section studies cross-section regression in time compressed data. Let $z_{it}=(x_{it},y_{it})^\prime$  be  a vector of $n$ variables  characterized by $B_0 z_{it} =B_1 z_{i,t-1}+\epsilon_{it}$, where $B_0$ and $B_1$ are the same across units. For example, when  $n=2$,  
\[ B_0=\begin{pmatrix} 1 & -\kappa_0 \\ -\beta_0 & 1 \end{pmatrix}, \quad B_1=\begin{pmatrix} \rho & \kappa_1 \\ \beta_1 & \alpha \end{pmatrix}, \quad \epsilon_{it}=\begin{pmatrix} u_{it} \\ e_{it}\end{pmatrix}.
\]
 We start with the simple case when $\kappa_0=\kappa_1=\alpha$=$\beta_1$=0 so that letting $\beta_0=\beta$, 
\begin{align}
x_{it}&= \rho x_{it-1}+u_{it}\quad &\label{eq:eqx} \\
y_{it}&=\beta x_{it}+e_{it} ,\quad & 
\label{eq:eqy}
\end{align}
where $u_{it}\sim (0,\sigma^2_i)$ and  $e_{it}\sim (0,\tau^2_i)$ are  stationary and uncorrelated. When $\rho<1$, $x$ is stationary (or I(0)), and when $\rho=1$, $x$ is I(1).
 If $x_{it}$ is I(1) and $e_{it}$ is stationary,   $(1,-\beta)$ is a cointegrating vector in the sense of  \citet{engle-granger-87}. A time  series regression using $T$ observations of $y$ and $x$ of a particular unit $i$ will converge at the fast rate of $T$, but  the asymptotic distribution of $\hat\beta$ is, in general, non-standard.  If we focus on the low frequency variations as in \citet{mueller-watson:17}, we have a small sample problem because few cycles with long periodicity will be observed over a short span. Panel regressions also have a fast convergence rate of $\sqrt{N}T$, and estimation can be quite involved if there is cross-section dependence.  Instead, we consider least-squares estimation of a `timeless',  cross-section  between-group   regression  in time compressed data $\tilde Y_{iK}$ and $\tilde X_{iK}$:
\begin{equation}
 \tilde Y_{iK} =c+  \beta \tilde X_{iK} + \tilde e_{ik}.  
\label{eq:S}
\end{equation}  

We first consider the properties of $\hat\beta$ under the following  assumptions.

\paragraph{Assumption B:} (i) $u_{is}\perp e_{it}$ for all $s,t$ and for all $i$; (ii) $u_{it}\sim (0,\sigma^2_i)$ is  serially uncorrelated  with  $\mathbb E_i[\sigma^2_i]>0$; (iii) $e_{it}\sim (0,\tau_i^2)$ is stationary; (iv) $x_{i0}=0$ for all $i$;   (v) $\{x_{it},e_{it}\}$ and $(\bar X_{iK}, \bar e_{iK})$ are i.i.d. across $i$ with finite fourth moments.

\paragraph{Assumption  C: }(i) $\Phi=\lim_{K\rightarrow\infty} \Phi_K/K=\tilde a_K>0$ and  (ii) $A_0=\sum_{t=T_0+1}^{T_1} w_t^2=O(K^{-1})$.

\medskip

We assume that $u_{is}$ and $e_{it}$ are independent across $i$, and for all $s,t$ so that strict exogeneity holds.  We do not require $\sigma^2_i>0$ for every $i$, only that  $\mathbb E_i[\sigma^2_i]>0$. The condition  is needed for identification as it ensures  $Q_{XX}=\var_i(\tilde X_{iK})$ is non-degenerate.
Assumption B.(iii) allows $e_{it}$ to be serially correlated but cannot be non-stationary to rule out spurious regressions. Cross-section dependence is not allowed for now and Assumption B.(iv) simplifies the analysis. Assumptions B.(v)  allows us to use LLN and CLT for iid variables. In particular, $\frac{1}{N}\sum_i f(\bar X_{iK},\bar e_{iK})\pconv \mathbb E_i[f(\bar X_{iK},\bar e_{iK})]$, giving
\[ \frac{1}{N}\sum_{i=1}^N (\tilde X_{iK}-\bar {\tilde X}_{K})^2 \pconv \var_i(\tilde X_{iK})=\mathbb E_i[\sigma^2_i]\Phi_K\equiv Q_{XX},\]
and 
\[\frac{1}{\sqrt{N}} \sum_i (\tilde X_{iK}-\bar {\tilde X}_K)\tilde e_{iK}\dconv N(0,V_K), \quad \text{where } \quad V_K=\mathbb E_i[(\tilde X_{iK}-\mu_X)^2\tilde e_{iK}^2].\]  

The properties of the least squares estimation is determined by
\begin{eqnarray*}
(\hat\beta-\beta) &=&\frac{ N^{-1}\sum_{i=1}^N (\tilde X_{iK}-\bar{\tilde X}_K) \tilde e_{iK}} {N^{-1}\sum_i (\tilde X_{iK}-\bar {\tilde X}_K)^2}.
 \end{eqnarray*}

Under Assumption B, strict exogeneity ensures that the estimator is $\sqrt{N}$ consistent when  $x$ is  I(0)   as the denominator of  $\hat\beta-\beta$ converges to $Q_{XX}$ which is $O(1)$, and   the numerator scaled by $\sqrt{N}$ obeys a central limit theorem. But  faster convergence rates can be obtained, and Assumption C makes these requirements precise.  When $x$ is I(1),  we have a {\em  signal magnification} effect that is   scheme specific if C.(i) is satisfied since $\var(\tilde X_{iK})=\sigma^2_i \tilde a_K(w)K$.  Now $\var(\tilde e_{iK}) =\var(\sum_t w(t) e_{it})=\tau_i^2\sum_t w(t)^2= \tau_i^2 A_0(w)$.  If the time compression scheme is such that $\var(\tilde e_{iK})=O(K^{-1})$ as indicated in C.(ii), we have {\em noise dilution}.  Obviously, if we point sample at $t=T_1+1$ and $T_1$ is small, then $\Phi_K=1$ and  both C.(i) and C.(ii) would fail.
 If  we  point sample  at a large  $T_1$,  C.(i) is satisfied, but C.(ii) fails because are concentrated to one point, and  $\sum_t w(t)^2=1$.   Time compression schemes with spread out weights  satisfy both C.(i) and C.(ii), and of these,  $K$-averaging is the simplest.\footnote{To check C.(ii) in \textsc{matlab}, compute \texttt{sum(w.\textasciicircum 2)} which should be $O(1/K)$. To check C.(i), compute  \texttt{sum(cumsum(w.\textasciicircum 2))} which should be $O(K)$.}

\begin{proposition}
\label{prop:prop1}
Let $\hat\beta$ be estimated from a correctly specified static regression $\tilde Y_{iK}=c+\beta \tilde X_{iK}+\tilde e_{iK}$ where $(\tilde X_{iK},\tilde Y_{iK})=(\sum_{t=T_0+1}^{T_1}  w(t) x_{it}$, $\tilde Y_{iK}=\sum_{t=T_0+1}^{T_1}  w(t) y_{it})$.      Assume $\var(x_{i,T_0})=o(1)$. 
\begin{itemize}
\item[a.] Let  Assumptions A and B hold.  As $N\rightarrow\infty$  with  $K$ fixed and whether $x$ is I(1) or I(0),   (i)
  $\hat\beta\pconv \beta$  and   (ii) $\sqrt{N}(\hat\beta-\beta)\dconv N(0, Q_{XX}^{-1} V Q_{X}^{-1})$,    where $V=\var_i((\tilde X_{iK}-\mu_X)\tilde e_{iK})$. 

\item[b]  Let Assumption A, B and C.(i)  hold and $x$ is I(1). Let  $(\tilde X_{iK},\tilde Y_{iK})$. For large $K$,
$\sqrt{NK}(\hat\beta-\beta) \dconv N(0,\mathcal Q_{XX}^{-1} \mathcal V \mathcal Q_{XX}^{-1})$, $\mathcal V=\lim_{K\rightarrow\infty} V$,    $\mathcal Q_{XX}=\lim_{K\rightarrow\infty} K^{-1} Q_{XX}$.
%\textcolor{red}{Matrix $V$ is already a limiting variance in part (a). Maybe there is a better notation in parts (b), (c) and (d).}

\item[c] Let Assumption A, B, and C.(ii)  hold and $x$ is I(1). Then for large $K$, \newline
$\sqrt{NK}(\hat\beta-\beta)\dconv N(0,\mathcal Q_{XX}^{-1}  \mathcal V_1 \mathcal Q_{XX}^{-1})$, $\mathcal Q_{XX}=\lim_{K\rightarrow\infty} Q_{XX}$ and $\mathcal V_1= \lim_{K\rightarrow\infty} K^{-1} V$.

\item[d] Let A, B, and  C hold and $x$ is I(1). Let  $(\tilde X_{iK},\tilde Y_{iK})=(\bar X_{iK},\bar Y_{iK})$  under K-averaging.
 For large $K$,     $\sqrt{N}K(\hat\beta-\beta) \dconv N(0,\mathcal Q_{XX}^{-1} \mathcal V \mathcal Q_{XX}^{-1})$, $\mathcal V=\lim_{K\rightarrow\infty} V$, $\mathcal Q= \lim_{K\rightarrow\infty} K^{-1} Q_{XX}$.

\end{itemize}

\end{proposition}
 The statement on $\var(x_{i,T_0})$ makes clear that limiting variation is due to a large $K$ and not a large $T_{init}$. 
As stated in part (a),  $\hat\beta$ is $\sqrt{N}$ consistent and asymptotically normal for fixed $K$ regardless of whether $x$ is I(1) or I(0). Consistency follows from the fact that $u_{it}\perp e_{is}$ by assumption, and $\cov_i(\tilde X_{iK},\tilde e_{iK})=0$ for an $K$. If $K=1$,  $K$-averaging and point sampling reduce to  the standard cross-section result that $\hat\beta$ is $\sqrt{N}$ consistent and asymptotically normal, even when $x$ is I(1).

 Recall that $\var(\tilde e_{iK})= \tau_i^2 A_0(w)$  and $\mathbb E_i[(X_i-\mu_X)^2]=Q_{XX}=\mathbb E_i[\sigma^2_i]\Psi_K$. 
Under  homoskedasticity, the asymptotic variance of $\hat\beta$ is simply $\frac{\tau^2}{\sigma^2} \frac{A_0(w)}{ \Psi_K}$ when $x$ is I(0).   The quantity $\frac{\tau^2 }{\sigma^2}$ is the noise in the model relative to the signal in the data.  But from Lemma \ref{lemma:smooth},  $\Psi_K\rightarrow \Phi_K=\tilde a_K(w) K$ as $\rho\rightarrow 1$. Thus, when $x$ is I(1) or is highly persistent,  
\begin{equation}
\label{eq:eqVhomo}
 \sqrt{N}(\hat\beta-\beta) \dconv N\bigg(0,\frac{\tau^2}{\sigma^2} \frac{A_0(w)}{ \tilde a_K(w) K}\bigg).
\end{equation}
Because the cross-section variance fans out,   a faster convergence rate is now possible when $K$ is large.  The precise  rate depends on how fast  the variance of $\hat\beta$ tends to zero and is scheme specific through   $w(t)$.
  If there is signal magnification but no noise dilution as in point sampling at large $K=T_1$, then as  indicated in part (b), $\hat\beta$ is  $\sqrt{NK}$ consistent. It is possible for $x$ to be I(1) and yet there is no signal magnification. As indicated in part (c),  the estimator is $\sqrt{NK}$ consistent provided there is noise dilution. If both conditions of Assumption C are satisfied, the estimator  is $\sqrt{N}K$ consistent. The fastest convergence rate obtains when $K=T$.   
%$\sqrt{N}(\hat\beta-\beta)\dconv N(0, \frac{\mathbb E_i[x_{i,T_1}\tau_i^2]}{(\mathbb E_i[ x_{i,T_1+1}^2]^2)})$. 

 Proposition \ref{prop:prop1} assumes that the cross-section variance of  $x$  fans out at a rate of $K$, but  the rate can be slower or faster under alternative assumptions, as discussed above.  
In the case of an exogenous drift $g_i$ for example,   $\bar X_{iK}-\mu_X= \frac{K}{2}(g_i-\mathbb E_i(g))+\sum_{s=T_0+1}^t u_{is}=O(K)$, $\bar e_{iK}=O(K^{-1/2})$.  The meat of the sandwich variance is 
  $V_K=O(K)$, and the asymptotic sandwich variance is $O(K^{-4})O(K)=O(K^{-3})$. It follows that $\hat\beta$ from  the static regression in K-averaged data is  $\sqrt{N}K^{3/2}$ consistent and asymptotically normal, faster than the $\sqrt{N}K$ rate  when $g_i=0$.\footnote{The result parallels the convergence rate of $T^{3/2}$ in a time series regression of $x$  on  a linear trend.  However, if $\var_i(g_i)=0$, the only source of fanning out is  the stochastic trend and the convergence rate remains $\sqrt{N}K$.
If $\mathbb T_1>K$, $\var(\bar X_{iK}) =\mathbb T_1 \mathbb E_i[\sigma^2_i]$.}
Another example is when $x_{i,T_{init}}=0$,   then with  $\mathbb T_1=T_{init}+T_1$,  point sampling gives $Q_{XX}=\mathbb T_1 \mathbb E_i[\sigma^2_i]$  and $ V_K\approx \frac{\mathbb T_1}{K} \mathbb E_i[\sigma^2_i\tau_i^2].$
 Thus, $Q_{XX}^{-1} V_K Q_{XX}^{-1}=O(\mathbb T_1^{-1})$, giving  $\sqrt{N\mathbb T_1}$ consistent and asymptotically normal estimator as $\mathbb T_1\rightarrow\infty$ with $K$ fixed. When the fanning out rate is slower, the fastest convergence will not be  attainable, but it cannot be slower than $\sqrt{N}$  provided that $\cov_i(\tilde X_{iK},\tilde e_{iK})$ is of lower order than that of $\var_i(\tilde X_{iK})$.

In spite of the possibility of many convergence rates,  knowledge of the rate is not needed for inference, nor is there a need for HAC standard errors.  The reason  is that the scores $(\tilde X_{iK}-\mu_X)\tilde e_{iK}$ are iid across $i$. Hence a CLT for iid data applies in spite  of serial correlation structure of $e_{it}$, which is already  absorbed in $V$.  Asymptotic normality implies that  the usual $t$ statistic can be used for inference.   It would be a simple Wald test to see if the estimates in  non-overlapping samples are equal.  Nonetheless,  Assumption B.(iii) assumes  $u_{it}\sim (0,\sigma^2)$ is serially uncorrelated. With serial correlation, $\sigma^2_i$ would be replaced by $\omega^2_i$.   In the  IMA(1) case,  $\omega^2_i=(1+\vartheta_i)^2\sigma_i^2$.  If $\vartheta_i=-1+\frac{\bar\vartheta_i}{\sqrt{T}}$ as in  the   `nearly integrated, nearly white noise' case considered in \citet{perron-ng-restud}, then $\omega_i^2 \tilde a_K(w)K = \bar \vartheta_i^2 \tilde a_K$. With $Q_{XX}=\mathbb E_i[\omega^2_i] \tilde a_K K=\mathbb E_i[\bar\vartheta_i^2]\tilde a_K$,  there will be no fanning out at all, and a $\sqrt{N}$ consistent estimate is all that can be achieved if there is no noise dilution. However, with noise dilution,  a $\sqrt{NK}$ rate can be attained as indicated in part (c) of Proposition \ref{prop:prop1}.

The LD is not in the class of functions considered, but some observations  can be made. The LD is similar to the filter used in \citet{kuznets:58} to suggest evidence of a 20-year economic cycle.  \citet{howrey:68}  noted that  the ten-year difference of two five-year moving-averages   could spuriously induce  a cycle with a 20 year periodicity even  if the raw data were white noise. Use of long-differenced data in time series regressions is also known to suffer from small sample biases, \citet{campbell-jef:01}. However,  the LD transformation is  used only once at $t=T_a$ and $T_b$ to  collapse the time variations into $\breve X_{iK}$.  The LD regression error is  $\breve e_{iK}= \bar e_{iM}(T_b)- \bar e_{iM}(T_a)$. While  noise is diluted at rate $M$,   $\var_i(\breve e_{iK})=2\tau_i^2/M$ is inflated  because of differencing.   Since $\var(\breve X_{iK})\approx \sigma^2_i (K-M/3)$ for LD, it has  sandwich variance  $\frac{2\tau^2}{M} Q_{XX}^{-1}$ under homoskedasticity. Compared to  $K$-averaging which has   $\var_i(\bar e_{iK})=\tau_i^2/K$ and $\var(\bar X_{iK})=\sigma^2_i K/3$,
\begin{eqnarray*}
\frac{\var(\text{LD})}{\var(\text{KA})}= \frac{2K(T_0+a_K)}{ M(K-M/3)}\approx \frac{2(3T_0+K)}{3M(1-M/(3K))}\approx\frac{2K}{3M}+\frac{2T_0}{M}. 
\end{eqnarray*}
This ratio only depends on the tuning parameters and   not   on $\sigma^2_i$ or $\tau^2_i$.    K-averaging should have a smaller variance when $\frac{K}{M}>\frac{3}{2}$. 

\begin{figure}[ht]
\caption{Density of $\frac{\hat\beta-\beta}{\hat\sigma_{\hat\beta}}$}
\label{fig:thm1}
\begin{center}
\includegraphics[width=6.5in,height=2.0in]{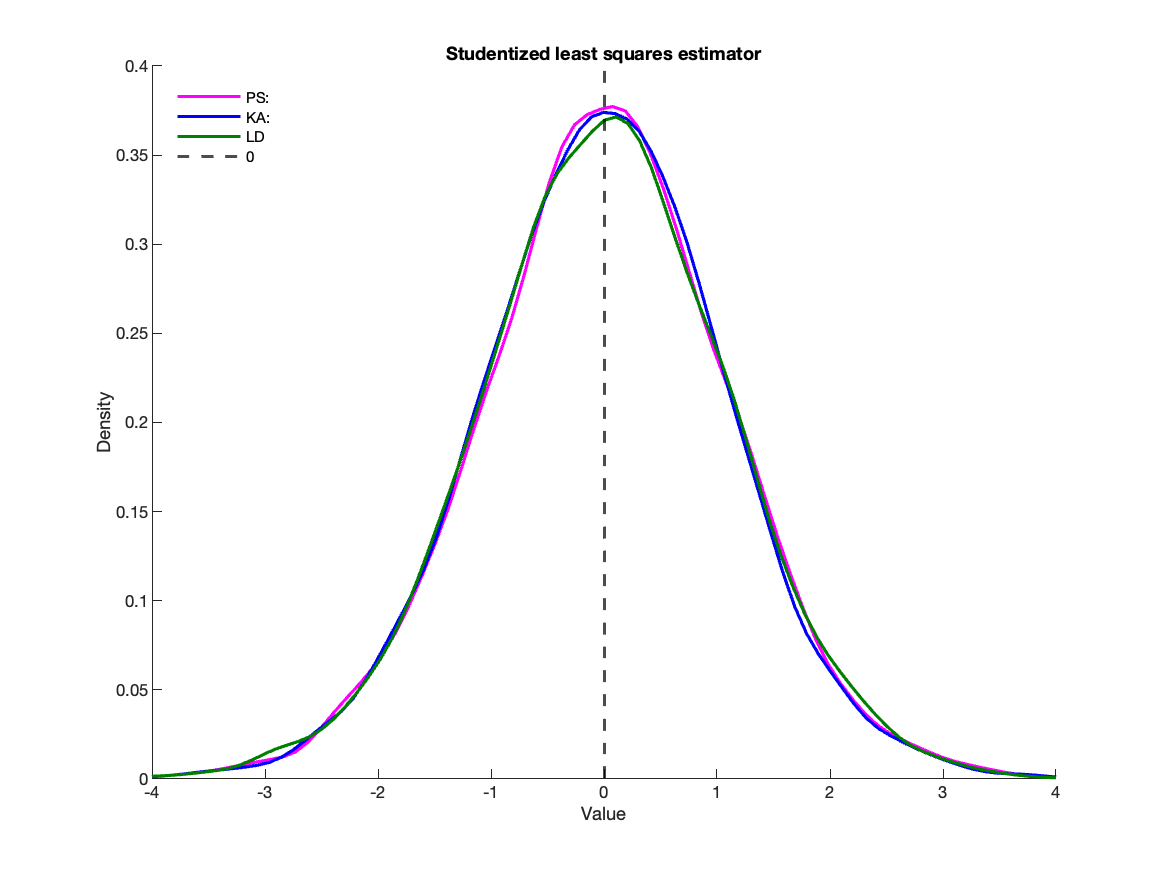}

\end{center}
\end{figure}

 To verify the theoretical predictions, we simulate data as 
\begin{eqnarray*}
y_{it}&=&\beta x_{it}  + e_{it}\\
x_{it}&=& x_{it-1}+v_{it}+\vartheta v_{i,t-1},
\end{eqnarray*}
  where  $(e_{it}, v_{it})$ are mutually uncorrelated Gaussian  random variables with  $(\sigma_e,\sigma_v)=( 5, 0.5)$.     We consider full sample averaging and point sampling at $T$. To make maximum use of the data with  $(T,N)=(50,50)$,   we use   $T_b=T-m-1, T_a=m+1$ for LD.   
Figure \ref{fig:thm1} plots the density of  $ (\hat\beta-\beta_0)$ standardized by White's standard errors for the case $\vartheta=0$ and  $\beta_0=-1$. The PS (in red), KA (in blue), and LD (in green) are all well approximated by the standard normal distribution. In spite of the similarity in the normalized distribution, there are distinctive differences in standard errors, being 0.014, 0.001, and 0.006 for PS, KA, and LD, respectively. The convergence rate of each estimator is determined by the rate at which its variance tends to zero. For this example, the ratios of the standard error at $T=50$ to $T=400$ are  7.975, 63.139, and 8.899, respectively. The ratios at $T=100$ to $T=400$ become 3.979, 15.833, and 4.206, respectively.  For PS, the ratio of standard errors is comparable to the ratio of the sample size, but for K-averaging, the ratio of standard errors is comparable to the ratio of the sample size squared. These ratios and the ones for other sample size suggest that the estimator using K-averaged converges  faster than point sampled data in the order of $\sqrt{T}$, supporting  the predictions of Proposition \ref{prop:prop1} that PS is $\sqrt{NT}$ consistent while K-averaging is $\sqrt{N}T$ consistent when $K=T$. Though our theorem does not include LD, the standard errors suggest that its convergence rate is also $\sqrt{NT}$, the same as PS.

\begin{table}[ht!]

\caption{Estimates and Rejection Rates, $(T,N)=(50,50)$}
% do-table2.m

\label{tbl:table2}
\begin{center}

  DGP: $y_{it}= \beta x_{it} + e_{it}$,   $e_{it}\sim N(0,3^2)$.

$x_{it}=x_{it-1}+v_{it}+\vartheta v_{it-1}$,   $v_{it}\sim N(0,0.5^2)$.

Regression: $\tilde Y_{iK}= c+\beta \tilde X_{iK}+\tilde e_{iK}$

\bigskip

\begin{adjustbox}{width=\textwidth}
\begin{tabular}{lr|rrr|rrrr|rrr}
 $\beta$ &  $\vartheta$  & \multicolumn{3}{c}{PS($T_1$)}  & \multicolumn{4}{c}{KA($T_0,T_1)$}  &   \multicolumn{3}{c}{LD$(m,T_0+m,T_1-m)$}\\ \hline
& &      $T_1=10$ & $T_1=20$ & $T_1=T$ & $(0,T)$ & $(10,20)$ & $(10,40)$ & $(20,40)$  & $m=0$ & $m=2$ & $m=5$\\
\hline
&& \multicolumn{5}{c}{Estimates of $\beta$} \\ \hline
\input table_thm1_est.tex  \hline
&& \multicolumn{7}{c}{Rejection Rates of $H_0: \beta=-1$} \\ \hline
\input table_thm1_tstat.tex \hline

\hline
\end{tabular}
\end{adjustbox} 
\end{center}
%{\footnotesize   $(e_{it},v_{it})\sim N((0,0),(3, 0.5)^2)$.} 
\end{table}

Next, we study the properties of $\hat\beta$  for different configurations of $(\beta,\vartheta)$. As seen  in the top panel of Table \ref{tbl:table2},  all  time-compressed data precisely estimate  $\beta$ for both values of $\vartheta$.
We then vary the true value of $\beta_0$ to evaluate the size of  testing $H_0: \beta^0=-1$.     The rejection rates are a bit over  5\%, but  the normal approximation of the $t$ statistic is reasonably accurate. Power is evaluated at  alternatives close to the null value of  $\beta^0=-1$. We see from the bottom panel that the power based on K-averaging is much higher than PS and LD. At the alternative of -0.95 (and $\vartheta=0$), for example, K-averaging has a rejection rate of 0.428, compared to a rate of 0.101 for PS and 0.128 for LD, due to their larger standard errors.

%, with the implication that $\sum \psi(\rho_i)^2\to\sum_t w(t)^2=O(K^{-1})$ when $\rho_i= 0$, and $ \sum \phi_s^2 =O(K)$ when $\rho_i= 1$. 

\section{More General Specifications}

Proposition \ref{prop:prop1} is built on  Assumptions A, B and C.    The  `meat' of the sandwich variance of the regression in time compressed data is  then
\begin{equation}
\label{eq:eqV}
 V=\mathbb E_i[ (X_i-\mu_X)^2 \tilde e_{iK}^{2}]= \mathbb E_i[(X_i-\mu_X)^2] \mathbb E_i[\tilde e_{iK}^{2}]+    \cov_i((\tilde X_{iK}-\mu_X)^2,\tilde e^{{2}}_{iK}).
\end{equation}
The covariance term in $V$ does not appear if  the model is correctly specified. However,
if this is not the case,  $\tilde e_{iK}$ will be replaced by $\tilde e_{iK}^*$, and  $\cov_i(\tilde X_{iK},\tilde e^*_{iK})$ may not be zero. The next three subsections   
 consider  cases when $\cov_i(\bar X_{iK},\bar e^*_{iK})\ne 0$ but super-consistency continues to hold.  We then consider a  case
when $\cov_i(\bar X_{iK},\bar e^*_{iK})=0$ and yet the variance of  $\bar e_{iK}^*$ precludes  the fastest  convergence rate.

\subsection{Relaxing  Strict Exogeneity}

 We assume    that the DGP is the static model  $y_{it}=\beta x_{it}+e_{it}$.   Assumption B.(i) asserts  $\cov_i(\tilde X_{iK},\tilde e_{iK})= \sum_{s}\phi_s \sum_t w(t) \mathbb E_i[u_{it}e_{is}]=0.$ 
We now  allow   $\text{cor}(u_{it},e_{is})=
 \gamma_{i,t-s}$ with $\sum_h |\gamma_{i,h}|< \infty$. Contemporaneous correlation occurs when $\gamma_{0,i}\ne 0$. When $\gamma_{h,i} \ne 0$ for $h<0$, $u_{it}$ is correlated with future $e_{is}$, while $h>0$ allows for feedback from   past values of $e_{it}$ to $u_{it}$.  

 Now $\tilde X_{iK}=x_{i,T_0}+\sum_s \phi_s u_{is}$ and $\tilde e_{iK}=\sum_t w(t) e_{it}$. Using
$\cov(u_{it},e_{is})=\gamma_{i,t-s} \sigma_i\tau_i$ for any  $i$,  
\begin{eqnarray*}
 \cov(\tilde X_{iK},\tilde e_{iK})&=&\cov\bigg( x_{i,T_0}+ \sum_{s=T_0+1}^{T_1} \phi_s  u_{is}, \; 
\sum_{t=T_0+1}^{T_1}w(t) e_{it}\bigg)\\
&=&
\sigma_i\tau_i\sum_{s=T_0+1 }^{T_1} \sum_{t=T_0+1}^{T_1} \phi_s w(t) \gamma_{i,s-t}\\
&=& \sigma_i\tau_i \Lambda_K(w,i),
\end{eqnarray*}
% Note that $|B_h|\le \max_s |\phi_s|\sum_t |w(t)|=O(1)$.  Assuming $\sum_h |\gamma_{h,i}|<\infty$,
%\[ \sum_{h=-(K-1)}^{K-1} \gamma_{ih}B_h \rightarrow \Gamma_i B_\infty,\]
%the unweighted sum of cross-covariances. It remains to find $B_h$ which does not depend on $i$.
%Consider the case of $K$-averaging when $w(t)=1/K$ and $\phi_s=\frac{(T_1-s+1)}{K}$. Grouping by lag $h=s-t\ge 0$ gives
%\begin{eqnarray*}
% \sum_{s-t=h}\phi_s w(t) = \frac{1}{K} \sum_{\substack{s-t=h\\ T_0+1 \le s,t\le T_1}} \frac{T_1-s+1}{K}
%&=&\frac{1}{K}\sum_{\substack{j=h+1}}^K \frac{(K-j+1)}{K} \rightarrow \frac{1}{2}=b_K.
%\end{eqnarray*}
%A similar result  holds for $h<0$. The
 %$b_K$ plays the role of $a_K$  in $K$ averaging. Furthermore, $\sum_{s}\sum_t \gamma_{i,s-t}\pconv \sum_{j=-\infty}^\infty \gamma_j =\Gamma$, the long-run cross-covariance. 
 where $\Lambda_K(w,i)=\sum_{h=-(K-1)}^{K-1} \gamma_{ih} B_h(w)$ and $ B_h(w)=\sum_{s-t=h} \phi_s w(t)$.  The Appendix shows that
$\Lambda_K(w,i) \rightarrow \Gamma_i B_\infty(w), $
where 
 $\Gamma_i=\lim_{K\rightarrow\infty}\sum_{h=-(K-1)}^{K-1} \gamma_{ih}$ is the long-run cross-covariance, and  $B_\infty(w) =\lim_{K\rightarrow\infty} B_h(w)$. 
Then \[ \cov(\tilde X_{iK},\tilde e_{iK})\rightarrow \sigma_i\tau_i\Gamma_i B_\infty(w).\]
 For  $K$ averaging where $w(t)=1/K$ for all $t$,  $B_h(w)\rightarrow 1/2$.  For point sampling at large $T_1$ where $w(t)=1_{t=T_1}$, $B_h(w)\rightarrow 1$. 
  With a constant $\tilde b_K$  that depends on  the time compression scheme,
\[\mathbb E_i[\cov(\bar X_{iK},\bar e_{iK})]=\mathbb E_i[\sigma_i\tau_i\Gamma_i\tilde b_K]=O(1).\]
Though the average covariance   is $O(1)$ due to endogeneity, the  cross-section variance  is  still $O(K)$. In consequence,  endogeneity bias vanishes with $K$.    Allowing $\cov(u_{it},e_{is})\ne 0$ only inflates variance,  in stark contrast to the I(0) case when  the estimators would be inconsistent.

Strict exogeneity may also fail because  of feedback.  Consider the dynamic model 
\begin{eqnarray*}
 x_{it}&=& x_{it-1}+\kappa y_{it-1}+u_{it}\\
y_{it}&=& \beta_0 x_{it}+\beta_1  x_{it-1}+e_{it}.
\end{eqnarray*}
Then $x_{it}=(1+\kappa \beta_0) x_{it-1}+\kappa \beta_1 x_{it-2}+\kappa e_{it-1}+u_{it}$ is still nearly I(1) if $|\kappa(\beta_0+\beta_1)|$ is small enough to  keep  $1+\kappa(\beta_0+\beta_1)$  in the $1/T$ neighbourhood of unity.  The effective innovation  to $x_{it}$ is now $u_{it}^*=\kappa e_{it-1}+u_{it}$. This implies that  $\cov(u_{it}^*,e_{i,t-1})=\kappa \tau^2$.  A static regression  of $\tilde Y_{iK}$ on $\tilde X_{iK}$ now has  two  biases when $K$ is fixed: one from omitting $\Delta x_{iK}$, and one from violating of strict exogeneity.   But signal magnification makes  the biases asymptotically negligible.  In both cases when strict exogeneity is violated, we  have  a $\sqrt{N}K$ estimate 
of  $\beta$ from  a static regression using $K$ averaged data, and a $\sqrt{NT_1}$ consistent estimate using data point  sampled  at $T_1$.

\subsection{Omitting Stationary Dynamics}
In time series regressions when $x$ is I(1), the bias from  omitting stationary dynamics may also be of second order. We now show that this is the case in cross-section regressions.
 Suppose  that  $x_{it}=x_{it-1}+u_{it}$ is strictly exogenous, but
\begin{equation}
y_{it}=\alpha y_{it-1} +\beta_0 x_{it}+\beta_1 x_{it-1}+ e_{it}. 
\label{eq:eqy1}
\end{equation}
 Following  \citet{bewley:79}, (\ref{eq:eqy1}) can be rearranged as
\begin{eqnarray*}
(1-\alpha) y_{it}&=& -\alpha \Delta y_{it}+ (\beta_0+\beta_1) x_{it}- \beta_1 \Delta x_{it}+e_{it}\\
y_{it} &=& \bigg(\frac{\beta_0+\beta_1}{1-\alpha}\bigg)  x_{it}-\bigg(\frac{\alpha}{1-\alpha} \Delta y_{it} +\frac{\beta_1}{1-\alpha}  \Delta x_{it}\bigg)+\frac{e_{it}}{1-\alpha}\\
&=& \beta x_{it}+ e_{it}^*,
\end{eqnarray*}
where  $\beta$,  often referred to  as the long-run multiplier, is
\begin{equation}
\label{eq:LRM} \beta=\frac{\beta_0+\beta_1}{1-\alpha}. 
\end{equation}
Since $x$ is I(1), $\beta$ is also the cointegrating coefficient 
characterizing the long-run relation between $y$ and $x$. It  can be computed as  the ratio of two $h$ period impulse response (IRF)  to an innovation in $x$ as $h\rightarrow \infty$: 
\[ \beta=
 \frac{\lim_{h\rightarrow\infty}\mathbb E_t [\partial y_{,it+h}/\partial u_{it}]}
{\lim_{h\rightarrow\infty} \mathbb E_t [\partial x_{i,t+h}/\partial u_{it}]} =
\frac{\lim_{h\rightarrow\infty}\text{IRF}_{u\rightarrow y,h}}{\lim_{h\rightarrow\infty}\text{IRF}_{u\rightarrow x,h}}
\]
The above analysis assumes that $x_{it}$ is non-stationary and strictly exogenous. Suppose that $x_{it}=\rho x_{it-1}+\kappa y_{it-1}+u_{it}$ and let $z_{it}=(x_{it},y_{it})'$. 
%We $z_t=(x_t,y_t)'$, 
We have a  system  of the form $\mathbb B_0 z_{it}=\mathbb B_1 z_{it-1}+ \epsilon_{it}$. Let $\mathbb A=\mathbb B_0^{-1}\mathbb B_1$ and $\mathbb C(L)=(I_2-\mathbb A L)^{-1}\mathbb B_0^{-1}$.\footnote{In this ADL(1,1) model, $\mathbb C(1)=(\mathbb B_0-\mathbb B_1)^{-1}=\frac{1}{D} \begin{pmatrix} 1-\alpha & \gamma \\ \beta_0+\beta_1 & 1-\rho \end{pmatrix}$, $D=(1-\rho)(1-\alpha)-\gamma (\beta_0+\beta_1)$.} Then,
\[ z_t =\mathbb C(L) \begin{pmatrix} u_{it} \\ e_{it} \end{pmatrix}\]
where $\mathbb C=(\mathbb C_0+\mathbb C_1 L+ \ldots \mathbb C_h L^h+\ldots)$, $\mathbb C_h$ is a $2\times 2$ matrix with $\mathbb C_{ij,h}$ in the $(i,j)$-th entry.
 The system is non-stationary if the largest eigenvalue of $\mathbb A$ is unity. In  this case, 
\[ \text{IRF}_{u\rightarrow y,h}=\mathbb C_{21,h}, \quad \text{IRF}_{u\rightarrow x,h}=\mathbb C_{11,h}.\]

If the system is stable, $\mathbb C_{11,\infty}$ and $\mathbb C_{21,\infty}$ are necessarily zero, and we define the long-run multiplier in terms of cumulative impulse responses: $\mathbb C_{21}(1)=\lim_{h\rightarrow\infty} \sum_{k=0}^h= \mathbb C_{21,k}$, and
$\mathbb C_{11}(1)$ is similarly defined.  The system implies
\[ \beta=\frac{\beta_0+\beta_1}{1-\alpha} = \begin{cases} \mathbb C_{21}(\infty)/\mathbb C_{11}(\infty)\quad  & \text{eig}_{\text{max}}(\mathbb A)=1 \\
\\
\mathbb C_{21}(1)/\mathbb C_{11}(1) \quad & \text{eig}_{\text{max}}(\mathbb A)<1\end{cases}.
\]
 Though $\mathbb C_{ij,h}$ depends on all parameters in the system, $\beta$ only depends on the parameters of the equation for $y$ because the other parameters enter $\mathbb C_{21}$ and $\mathbb C_{11}$ through a common determinant that gets canceled out. 

As $\beta$ is the long-run multiplier  whether or not the system is stable, and whether $x$ is strictly exogenous,     estimation of it is of interest. Many methods are available if we take a stand on whether the data are I(1) or I(0). Can we estimate $\beta$ without taking such a  stand? A static regression of $y$ on $x$  that omits $\Delta y_{it}, \Delta x_{it-1}$ would have $\mathbb E_i[e^*_{it}x_{it}]\ne 0$. If $x_{it}$ were stationary, a time series or a panel  regression would yield inconsistent estimates. But if $x_{it}$ is I(1),  a time series regression would yield  a super ($T$) consistent estimate because  the omitted variable bias is of second order, but the distribution of $\hat\beta$ is non-standard. In terms of  K-averaged data,
\begin{eqnarray*}
 \bar Y_{iK} &=& \beta \bar X_{iK}-\bigg( \frac{\alpha}{1-\alpha} \Delta y_{iK} +\frac{\beta_1}{1-\alpha} \Delta x_{iK}\bigg) +  \frac{\bar e_{iK}}{1-\alpha}\\
&=& \beta \bar X_{iK}- \frac{\eta_{iK}}{1-\alpha}+\frac{\bar e_{iK}}{1-\alpha}
\end{eqnarray*}
where $\Delta y_{iK}=\frac{1}{K}(y_{i,T_1}-y_{i,T_0})$, $\Delta x_{iK}=\frac{1}{K}(x_{i,T_1}-x_{i,T_0})$, and $\eta_{iK}= \alpha \Delta y_{iK}+\beta_1 \Delta x_{iK}$. Importantly,  the omitted $\eta_{iK}$ is negligible when $K$ is large.
\begin{proposition}
\label{prop:lrm-I1}
Suppose that $x_{it}$ is I(1), $y_{it}$ is generated by the dynamic model given in (\ref{eq:eqy1}) with $|\alpha|<1$   and Assumption B holds.   Let $\hat\beta$ be obtained by estimation of (\ref{eq:S}) using $K$-averaged data.
\begin{itemize}
\item[i] As $N\rightarrow\infty$ with $K$ fixed, 
$ \sqrt{N}\bigg(\hat\beta-\beta+\frac{\cov_i(\bar X_{iK},\eta_{iK})}{\var_i(\bar X_{iK})}\bigg)\dconv N(0,Q_{XX}^{-1}W_K Q_{XX}^{-1})$, where  $W_K=\var_i\bigg((\bar X_{iK}-\mu_X) \bar e^*_{iK}\bigg)$ with $\bar e^*_{iK}=\frac{\bar e_{iK}-\eta_{iK}}{1-\alpha}$.
\item[ii] If $x$ is I(1), then for large $K$,
$\sqrt{N}K(\hat\beta-\beta) \dconv N(0, \mathcal Q_{XX}^{-1} \mathcal W \mathcal Q_{XX}^{-1} )$, where $\mathcal W=\lim_{K\rightarrow\infty }  W_K$ and $\mathcal Q_{XX}=\lim_{K\rightarrow\infty} K^{-1} Q_{XX}$.  
\end{itemize}

\end{proposition}
 In the present setting, 
  the omitted variable  bias for a given $K$ is $\text{bias}=\frac{\cov_i(\bar X_{iK},\bar e^*_{iK})}{\var_i(\bar X_{iK})}$.  Since $\bar X_{iK}$ only depends on $\{u_{it}\}$ and $u_{it}\perp e_{is}$ by assumption, bias can only arise through $\cov_i(\bar X_{iK},\eta_{iK})$. As $\bar X_{iK}=O_p(K^{1/2})$ and  $\eta_{iK}=O_p(K^{-1})$ in the stationary case,  $\cov_i(\bar X_{iK},\eta_{iK})=O(-3/2)$. Upon scaling by $\var_i(\bar X_{iK})=O(K^{-1})$, the bias is $O(K^{-1/2})$. In the I(1) case, $\eta_{iK}=O_p(K^{-1/2})$, but $\bar X_{iK}=O(K^{1/2})$, so $\cov(\eta_{iK},\eta_{iK}=O(1)$
But upon scaling by $\var_i(\bar X_{iK})=O(K)$, the  bias  is $O_p(K^{-1})$ that vanishes with $K$.  We thus have a $\sqrt{N}K$  consistent estimate of the long-run relation $\beta$ as $K$ increases.
  Though $\bar e^*_{iK}$ does not contribute to the bias,   $\var(\bar e_{iK}^*)>\var(\bar e_{iK})$. Omitting the lags yields a less efficient estimate  of $\beta$  as reflected in $W_K$.

It should be noted that $\beta$  is distinct from the long-run effect of a unit shock to $x$, which is
\[ \lim_{h\rightarrow\infty} \mathbb E_t\bigg[\frac{\partial y_{it+h}}{\partial u_{it}}\bigg] = \beta \times \lim_{h\rightarrow\infty}\mathbb E_t\bigg[ \frac{\partial x_{it+h}}{\partial u_{it}}\bigg].\]
If $\beta\ne 0$, the two coincide only if $x_{it}$ is a random walk with innovation variance of one, for in that case, $\mathbb E_t[\frac{\partial x_{i,t+h}}{\partial u_{it}}]=1$ for any $h\ge 0$. In other cases,  knowledge about the dynamics of $x_{i}$ is  necessary to pin down  the long-run effect.
%If $\Delta x_{it}=C_i(L) u_{it}$ and thus $ x_{it}=C_i(1) u_{it}+C_i^*(L)\Delta u_{it}$, then $\lim_{h\rightarrow\infty} \frac{\partial x_{it+h}}{\partial u_{it}}=C_i(1)$   because the second term is  dominated by the first.
For example, in the IMA(1) case when  $\Delta x_{it}= u_{it}+ \vartheta_i u_{it-1}$ and $u_{it}\sim (0,\sigma^2_i)$,
the  long-run effect of  $u_{it}$ is $\beta(1+\vartheta_i) \sigma_i\ne \beta$. 
 Consistent estimation of  $\beta$ (the long-run multiplier  of a change in the level of $x$) is  not enough to characterize the long-run effect of an innovation to $x$. We also cannot extrapolate from a relation estimated from variations of medium periodicity to a medium term effect of a shock to $x$. 
 Any $h$ period  effect necessitates  estimation of the system $\mathbb B_0 z_{it}=\mathbb B_1 z_{i,t-1}+\epsilon_{it}$, or by local projections.  Comparison of the long-run multiplier $\hat\beta$  with the impulse responses  such as constructed in \citet{bilal-kanzig:24} should be made with caution.

\subsection{Heterogeneous Coefficients} 
The above analysis shows that the strong cross-section variability generated by I(1) data  usually accelerates the convergence rate of  $\hat\beta$. We now consider a  counter example when the strong signal does not accelerate convergence.

  Consider  heterogeneous coefficient $\beta_i= \beta+ \zeta_i$, where  $\mathbb E_i[\zeta_i]=0$, and $\var_i(\zeta_i)=\sigma^2_\zeta$. The model $y_{it}=\beta_i x_{it}+e_{it}$ implies 
$ y_{it}=x_{it}\beta+ e_{it}^*$, where  $e_{it}^*= \zeta_i x_{it}+e_{it}$, so that
$ \cov(x_{it},e_{is}^*)=\mathbb E[\zeta_i x_{it} x_{is}]\ne 0.$
With  $\bar e_{iK}^*=\bar e_{iK}+\zeta_i\bar X_{iK}$,
a static regression in $K$-averaged data  yields
\[ \hat\beta = \beta
+ \frac{\cov_i(\zeta_i,\bar X_{iK})}{\var_i(\bar X_{iK})}=\beta+\frac{\mathbb E_i[ \zeta_i(\bar X_{iK}-\mu_X)^2]}{\var_i(\bar X_{iK})}.
\]
If $\zeta_i\perp(\sigma^2_i, x_{i,T_0})$, 
%\[\mathbb E_i[v_i(\bar X_{iK}-\mu_X)^2]=\mathbb E_i[v_i\bar X_{iK}^2]+ \mathbb E_i[v_i(x_{i,T_0}-\mu_X)^2]=0.\]
the orthogonality condition is satisfied because $\mathbb E_i[\zeta_i]=0$, but the estimator is only $\sqrt{N}$ consistent. This result is best understood in the homoskedastic case when $\sigma^2=\mathbb E_i[\sigma_i^2]$ and $\tau^2=\mathbb E_i[\tau_i^2]$. Then,   $\var(\bar e_{iK}^*)= \sigma^2_v \mathbb E_i[\bar X_{iK}^2]+\var(\bar e_{iK})=O(K)$. 
$K$-averaging does not achieve  noise dilution because parameter heterogeneity scales $\bar e_{iK}^*$  with $\bar X_{iK}$, and any signal magnified by time compression is offset by a magnified noise. Thus, even if $x$ is a random walk with drift, the rate will remain $\sqrt{N}$. \citet{pesaran-smith:95} considered a heterogeneous coefficient model where  $x$ is I(1) with a heterogeneous drift. They find that the coefficient estimate from a cross-section  regression  in full-sample averaged data is consistent. We show that the convergence rate is $\sqrt{N}$, whether or not there is a drift. It should be remarked that point sampling at $T_1$ will also be $\sqrt{N}$ consistent because it also has the  problem that $\tilde e_{i,T_1}=e_{i,T_1}+\zeta_i X_{i,T_1}$ scales with $X_{i,T_1}$.  The practical implication is that if a researcher wants to allow for heterogeneous coefficients, the usual rate of  $\sqrt{N}$ should be expected.

\section{Empirical Issues}
We have so far assumed that $x_{it}$ only has one component that is a stochastic trend.  In applications of empirical interest,  $x_{it}$ often has additional sources of variation. Subsection 1 considers fixed effects and cross-section dependence.  Subsection 2 considers the case when the two components are observed. Subsection 3 considers coexistence of latent permanent and transitory  components.

\subsection{Fixed Effects and Cross-Section Dependence}

A common critique of cross-section regressions is the bias from omitted  time-invariant  variables  and cross-section dependence.\footnote{See, for example, \citet{deschenes-greenstone:07}, \citet{mendelsohn-massetti:17}, and \citet{andrews-ecta-05}.}  To investigate these issues in our setup,  suppose that
\[ y_{it} =\beta x_{it}+ s_i f_t+e_{it},\]
where $x_{it}$ is I(1), $s_i$ is a unit-specific effect, $f_t$ is  time variation common across $i$.    This interactive fixed effect specification considered in \citet{bai-ecma:09} nests the additive fixed effect  as a special case.   Few panel data analyses can allow for both I(1) and I(0) common factors. In our setting, I(1) common factors must be controlled for to achieve super-consistent estimates;  otherwise, the omitted variations would be the same order as $x$. Then, as in the case of heterogeneous coefficients, the best that can be achieved is $\sqrt{N}$ consistent estimates. 
To make progress,  let $f_t=\mu_f + f^0_t$ be stationary and $s_i=\mathbb E_i[s_i]+s^0_i$, where $\mu_f=\mathbb E[f_t]$ with  $\mathbb E[ f^0_t]=\mathbb E_i[s^0_i]=0$. Because $f^0_t$ is stationary,
\begin{eqnarray*}
s_i\bar f_K
&=& \underbrace{\mathbb E_i[s_i]\mu_f}_{\text{constant}} + s_i^0 \mu_f+\underbrace{s_i \bar f^0_K}_{O(K^{-1/2})}.
\end{eqnarray*}
In this set up, the regression model in $K$-averaged data is
\begin{eqnarray*}
 \bar Y_{iK}&=&\beta \bar X_{iK}+  s_i\bar f_K + \bar e_{iK}\\
&=& c  + \beta \bar X_{iK} +s^0_i\mu_f + \bar e^*_{iK},
\end{eqnarray*}
where $c$ absorbs the constant term in $s_i\bar f_K$ and  $\bar e^*_{iK}=\bar e_{iK}+ s_i\bar f^0_K$.
The  omitted variable bias  is
  \[ \text{bias}=\frac{\cov_i(\bar X_{iK},s_i^0\mu_f)}{\var_i(\bar X_{iK})}=\frac{\mu_f \cov_i(x_{i,T_0},s_i^0)}{\var_i(\bar X_{iK})}.\]
  There is trivially no bias when $\mu_f=0$ and  $\hat\beta$ continues to be $\sqrt{N}K$ consistent.

More interesting is when $\mu_f\ne 0$,  the important special case being $f_t=\mu_f=1$ which makes $s_i$  an individual fixed effect. 
If  $x_{i0}=0$,  strict exogeneity ensures that $\cov_i(x_{i,T_0},s^0_i)=0$ as is required for consistent estimation of $\beta$.  The question is at what rate.  Now $\bar e_{iK}^*=\bar e_{iK}+s^0_i=O(1)$. While $\bar e_{iK}=O(K^{-1/2})$ and vanishes with $K$,  $s^0_i$ cannot be diluted away. So while K-averaging will enjoy some noise dilution for a given $K$, $s^0_i$ will eventually dominate and  inflate   $V$ in the  sandwich variance from $O(1)$ to $O(K)$.\footnote{The situation is different from  \citet{andrews-ecta-05}. In Andrews' case,  common shocks create   cross-section dependence and invalidate the usual standard errors for the $\sqrt{N}$ consistent estimate.} As a consequence,   $\hat\beta$ is  $\sqrt{NK}$  (instead of $\sqrt{N}K$) consistent. However,   the standard errors are valid if $s_i^0$ is independent across $i$. While the convergence rate using $K$-averaged data is downgraded upon losing the benefit of noise dilution, this is not the case if  we point sample at $T_1$ because it does not have the noise dilution property anyway. With $\tilde e_{i,T_1}=e_{i,T_1}+s^0_i$ and $\var_i(e_{i,T_1}^*)=O(1)$, the order is the same  as $\var(e_{i,T_1})$ when $s^0_i$ was absent, and the convergence  rate remains $\sqrt{NT_1}$. The LD differences away the fixed effect to  identify $\beta$ from the change rather than level of the non-stationary component, and the convergence rate should be $\sqrt{N}K$, a distinct advantage over $K$-averaging if the fixed effect is truly of the additive type.

The above arguments break down if  $s^0_i$ is correlated with   $x_{it}$, as strict exogeneity only rules out correlation between $u_{it}$ and $e_{is}$. However, this bias can be mitigated in different ways. First,  $s^0_i$ can be controlled by incorporating  any variable $z_i$ correlated with $s_i$.  
A second  possibility is to note that  the numerator of the bias  $\mu_f\cov(x_{i,T_0},s_i^0)=O(1)$  does not grow with $K$ or $T_0$, implying that the stronger is signal magnification, the faster is convergence. For example,  under the  $T_{\text{init}}=0$ setup,  $\var_i(\bar X_{i,T_1})=O(\mathbb K)$ where $\mathbb K=T_{\text{init}}+K/3$, and $\hat\beta$ is
%. . Hence for large $\mathbb T$ relative to $K$, $Q_{XX}=\var_i(\bar X_{iK})=O(\mathbb T)$. It follows that the bias is $O(\mathbb T^{-1})$ and vanishes as $\mathbb T\rightarrow \infty$.   Since $\bar X_{iK}-\mu_X=O(\mathbb T^{1/2})$ and $Q_{XX}=O(\mathbb T)$, we have $V_K=O(\mathbb T)$. The sandwich variance $Q_{XX}^{-1}V_K Q_{XX}=O(\mathbb T^{-1})$ implies a 
 $\sqrt{N\mathbb K}$ consistent. Similarly, an exogenous drift will also give an estimator with fast enough convergence to reduce the bias from the omitted fixed effect. When the noise dilution mechanism is active, such as in $K$-averaging,  omitted fixed effects and cross-section dependence impact the convergence rate of the estimator which  depends on the underlying profile of the cross-section variance, but it is at least $\sqrt{N}$ consistent if the model is otherwise well specified.
%A third possibility arises when  $x_{it}$ has a drift of $g_i$ with $\var_i(g_i)>0$. Then as noted in Remark 1, $Q_{XX}=O(K^2)$ and the bias is $O(K^{-2})\rightarrow 0$. With $\bar e_{iK}^*=O(1)$,  $V_K=O(K^2)\cdot O(1) $ , giving a sandwich variance is $O(K^{-2})$. The deterministic drift thus yields a convergence rate of $\sqrt{N}K$. Putting $K=T$ would yield  the most favourable rate for $K$-averaging when $s_i$ is correlated with the $x_{it}$,  

  To illustrate, consider a  regression of log consumption on log real GDP, both in per capita terms    using  data for N=44 countries from the Penn World Table from 1951 to 2023 for $T=73$  periods.  In a stochastic growth model, consumption and output  share a common stochastic trend associated with productivity, and their ratio  is one. But there may be other sources of variation which can only be learned from the data.
As benchmark, we consider three   panel estimators  for non-stationary data with cross-section dependence captured by common factors $F$ and $\Lambda$ so that  $y_{it}=x_{it}\beta+\lambda_i F_t+ e_{it}$. All three estimators have a theoretical convergence rate of $\sqrt{N}\cdot T$. The first is the LSDV, which  is pooled least squares estimation in demeaned data. The estimate of 0.927   is consistent if $x$ and $F$ are strictly exogenous but has an asymptotic bias otherwise.   The second is the two-step  estimator (2sFM) of \citet{bai-kao:06} assuming that  $\lambda_i$ and  $F_t$ are of dimension two.  By first estimating the two factors, a $\hat\beta$ that does not adjust for long-run covariance and  endogeneity from possible correlation between $F_t$ and $x_{it}$ is obtained.   The third is the continuously updated, bias-corrected (cupBC) estimator of \citet{bkn}. The estimate of 0.950 removes asymptotic bias caused by unit roots, serial correlation, and endogeneity.\footnote{Two other alternatives use cross-section average in observables to capture cross-section dependence. 
The CEE-ARDL estimator in \citet{chudik-pesaran-joe:15} requires the factors to be serially uncorrelated and uncorrelated with the regressors. The latter is CEE-DL due to \citet{chudik-etal:16}. With this approach,  the errors will be correlated with the regressors when there is feedback, and is sensitive to the stationarity assumption.}   The cupBC estimates are serially correlated, but only one of the 44 series has a  $\hat\rho_i$ estimated from the autoregressive model $\hat e_{it}=\rho_i \hat e_{i,t-1}+\epsilon_{it}$ that is very close to unity.

\begin{table}[ht!]
\caption{Great Ratio Regression of log Consumption on log GDP, $(T,N)=(73,44)$}
\label{tbl:tbl_app05}
\begin{center}
\begin{tabular}{r|rrrrrrr}
    & cupBC & LSDV & 2sFM & PS2 & KA1 & KA2 &  LD2 \\ \hline
  est & 0.950 &  0.927 &  0.949 &  0.862 &  0.906 &  0.935 &  0.787 \\
     se & 0.003 &  0.003 &  0.001 &  0.066 &  0.029 &  0.029 &  0.078 \\

\hline
&& \multicolumn{6}{c}{Simulations} \\ \hline
\input mc_app05_bkn_sample.tex

\hline
\end{tabular}
\end{center}
{\footnotesize Note: cupBC is the continuously updated, bias-corrected estimator of \citet{bkn}. LSDV is least squares estimation in demeaned data. 2sFM is the two-step estimator of \citet{bai-kao:06} that does not correct for endogeneity bias. PS, KA, and LD are cross-section regressions in  point-sampled, K-averaging, and long-differencing. KA1 only uses GDP as regressor. PS2, KA2, LD2 additionally use emp/pop to control for cross-section dependence. PS is evaluated at $T_1=2023$ and LD at $(m, m+1,T-m)$ with $m=2$.}
\end{table}

The first factor estimated from  cupBC is  cyclical and appears stationary, while the second appears to be flat until around 1990 when a trend  emerges. The trend may be due to the fact  that our sample includes developing countries which have seen  increased participation rate. Let $x_{1,it}$ be log GDP, $x_{2,it}= \text{emp}_{it}/\text{pop}_{it}$,  where emp and pop are employment and population data.  Let KA2 be estimates using  K-averaged data in a cross-section regression with $\bar X_{1,iK}$ and $\bar X_{2,iK}$ as regressors.
The full sample  KA2  estimate  of 0.935 is close  to cupBC of 0.950, while  the point sampled estimate (PS2) at $T$ is 0.862, and the LD2 estimate with $(m,T_a,T_b)=(2,m+1,T-m)$ is 0.787.  
  The standard error of KA2 is 9.66 times larger than that of cupBC, and $\sqrt{T}$ in this example is 8.54. This suggests that there are omitted fixed effects  that slowed convergence of the estimates using K-averaged data  from the fastest possible rate of $\sqrt{N}\cdot T$ to $\sqrt{NT}$, as noted above. 

\begin{figure}[!ht]
\caption{Regression of Log Consumption on Log GDP}
\label{fig:app05}
\begin{center}
\includegraphics[width=6.0in,height=2.0in]{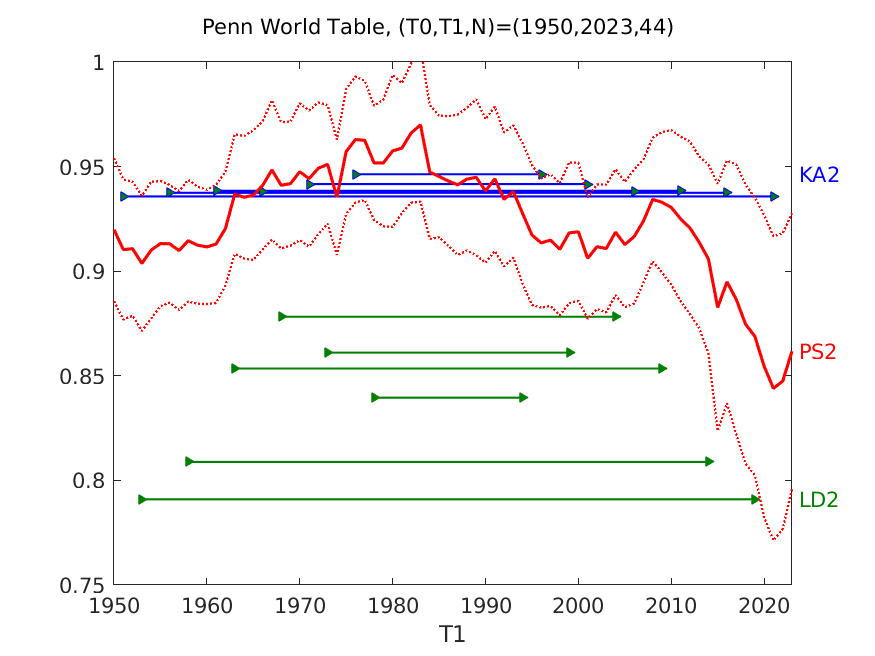}
\end{center}
{\footnotesize Note: The blue (K-averaged) and green (LD) lines are  estimates  for six  choices of ($T_0,T_1$) pairs. The red line is the estimate from point sampling at every $t$, and its one-standard error confidence interval in dotted red.}
\end{figure}

If the regression model is well-specified, $\hat\beta$  should  be fairly stable with respect to $K$, and  in our experience, the sensitivity of the K-averaging estimates  to $K$ is a useful diagnostic of the adequacy of the model. Without $\bar X_{2i,K}$,  the  KA1 estimates  are quite  sensitive to the choice of $T_0$ and $T_1$.  
 Figure \ref{fig:app05}   plots the  estimates   $(T_0,T_1)=(\kappa,T-\kappa)$  pairs for $\kappa$=2:5:30 and their standard errors.  We see that the KA2 estimates (in blue) are stable  with respect to $K=T_1-T_0$, but the point-sampled estimates  PS2 (in red)  display significant  variability  after 2000. Though within two standard errors of panel estimates,  PS2 at $T_1=2023$ is only 0.862. The   LD2 estimates are also sensitive to the choice of $T_0$ and $T_1$. With $m=2$, the LD2 estimate from the longest sample is 0.787,  though much lower, it is roughly within two standard errors of KA2 and the panel estimates.

To be confident of the estimates, a calibrated monte-carlo exercise with 10,000 replications is conducted by generating data using the cupBC estimate  of 0.950 with  the two estimated factors  fixed.  The corresponding  loadings,  the individual fixed effect, and  the autocorrelation coefficients $\hat\rho$ of the cupBC residuals are resampled.  For each $i$, synthetic AR1 errors are generated using $\hat\rho_i$ and  noise $\epsilon_{it} \sim N(0,\sigma^2_{\hat\epsilon,i}$). The results are shown in the bottom panel of Table \ref{tbl:tbl_app05}. The exercise shows that  the panel estimators are more efficient, as should be the case, because of explicit modeling of the common factors and biases from different sources. The cross-section estimates perform well considering that no correction is made, and standard normal inference is possible. The LD2 and PS2 estimates are closer to KA2 and the panel estimates than in the data. Further investigation reveals that if we only resample  $\Lambda_1$ but not $\Lambda_2$ (corresponding to the non-stationary factor), the mean PS2 and LD2 estimates in the monte carlo are close to the point estimates from the data.  This points to the sensitivity of both estimators to the choice of $T_0$ and $T_1$ for the data being analyzed, and more generally that differencing may not adequately control for common variations of the interactive type.

\subsection{Fanning-in vs Fanning-out}
\begin{figure}[!ht]
\caption{Share of Old and Young in Working Population}
\label{fig:young}
\includegraphics[width=7.0in,height=1.500in]{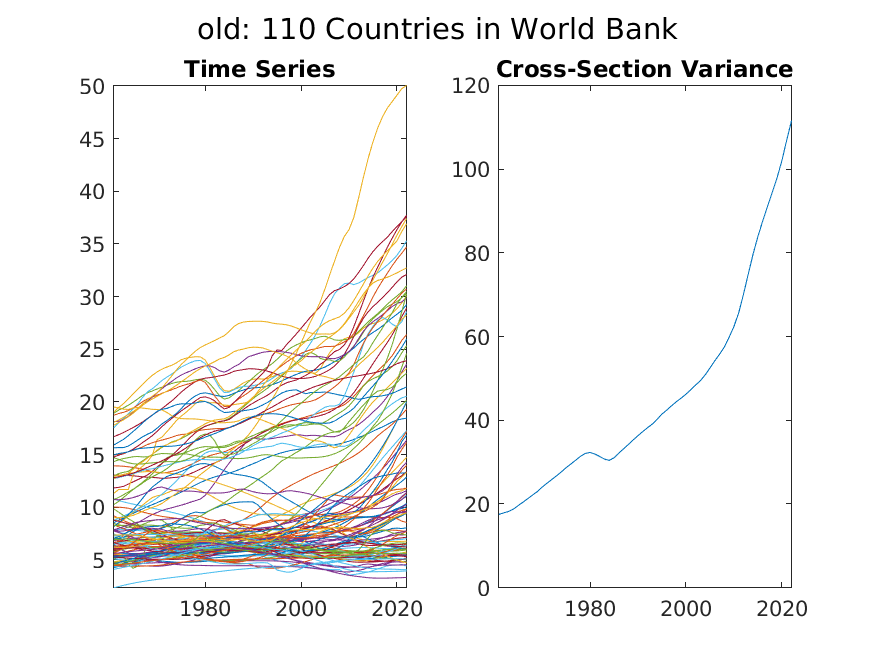}
\medskip
\includegraphics[width=7.0in,height=1.50in]{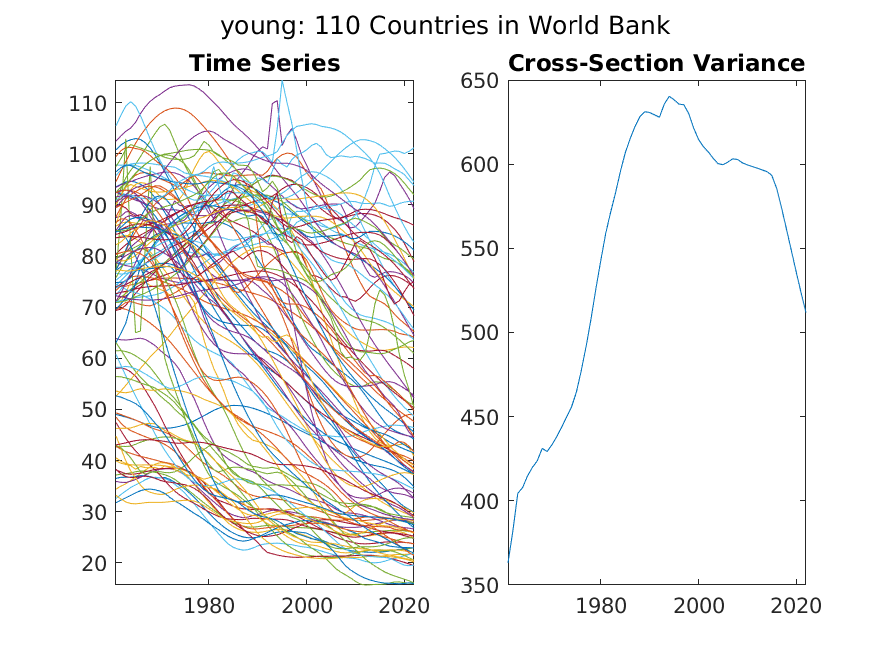}
\end{figure}

This subsection considers the case when $x$ is the sum of two observed variables with distinctive properties.
 Consider   the dependency ratio which  is the sum of  the population under 16 ('young') and over 65 ('old') relative to the working age population.  Using data from the 110 countries in the World Bank database from 1960-2022, the top panel of Figure \ref{fig:young} shows that `old'  is non-stationary;  the time series data have been steadily increasing, while its cross-section variance  fans out. The bottom panel shows that while the time series data of the `young' are also non-stationary, the cross-section variance  fans out until the turn of this century.  The dependency ratio and population growth exhibit a similar feature as `young'.  A possible explanation  is the  global decline  in fertility which creates  convergence across units. The cross-section variance of the dependency ratio  thus has a fanning-out component  (due to the 'old') and a fanning-in (due to the 'young')  component. 

 A high dependency ratio  has been suggested as a possible explanation for higher inflation and  a drag on growth.\footnote{See, for example, \citet{juselius-takats:21} for inflation. \citet{gordon:16} argues that an aging population creates a structural imbalance between high savings and low investment.  A different view is given in \citet{acemoglu-restrepo:17}.} The mixed evidence could be due to differences in the choice of the sample, as the relation between economic and demographic variables could be different for countries in different stages of economic development.  We use a sub-sample of 35 countries with  average real per-capita GDP  in the past 10 years over 10,000 USD, and with maximum inflation in the sample under 100 percent. In this data, the cross-section variance of the `young'  fans-in over the entire sample, distinctly different from the fanning out cross-section variance of `old'. If we use  the dependency ratio as regressor, we find that the estimate  is positive  in the inflation regression  and significant but varies significantly as $K$ increases. The coefficient in the  growth regression is more stable but is not significant. Because the dependency ratio has two components with different properties, we explore their distinctive economic impact.

\begin{table}
\caption{Growth/Inflation Regressions on Demographic Variables: K-averaging}
\label{tbl:demo}
\begin{center}
\begin{tabular}{ll|llll|llll}
 $T_0$ & $T_1$ & \multicolumn{4}{c}{inflation} & \multicolumn{4}{c}{growth} \\  \hline
&&  $\hat\beta_{old}$ & $t_{old}$ & $\hat\beta_{young}$ & $t_{young}$ 
&  $\hat\beta_{old}$ & $t_{old}$ & $\hat\beta_{young}$ & $t_{young}$ \\ \hline
\hline
 1963 &  2020 & 0.289 &  1.495 &  0.406 &  1.508  &-0.154 & -2.338 & -0.048 & -1.542 \\
 1968 &  2015 & 0.302 &  1.548 &  0.465 &  1.558 &-0.172 & -2.794 & -0.048 & -1.653 \\
 1973 &  2010 & 0.333 &  1.619 &  0.545 &  1.574 &-0.176 & -2.885 & -0.055 & -1.928 \\
 1978 &  2005 & 0.354 &  1.740 &  0.623 &  1.597 &-0.164 & -2.512 & -0.062 & -1.939 \\
 1983 &  2000 & 0.281 &  1.693 &  0.603 &  1.480 &-0.152 & -1.849 & -0.065 & -1.665 \\
\hline
\end{tabular}
\end{center}
{\footnotesize Regressions of Inflation (left) and Growth (right) on `old' and 'young' for a subsample of 35 countries, 1960-2022. The 'old' and `young'  are from the World Bank database, and the economic variables are from the Penn World Table.}
\end{table}

Table \ref{tbl:demo} reports results from regressions of inflation (left) and growth (right) on the demographic variables for  $(T_0, T_1)=(\kappa,T-\kappa)$ pairs with $\kappa$=2:5:25.
% The top panel  plots the  estimated coefficient  on the dependency ratio as $K$ (or $T_1$) increases, with $T_0=0$, along with the one standard error of the estimates. The estimate  in the inflation regression is positive and significant but varies significantly with $K$. The coefficient in the  growth regression is more stable but is not significant.  We then  enter  `old' and `young' as two separate regressors.  
For inflation, the full sample estimate of the coefficient on `old' is 0.289, and  the coefficient on `young' is 0.406, but neither is statistically significant at the 10\% level. For the growth regression, the full sample  estimate of the coefficient on 'old is -0.154 with a $t$ statistic of -2.338 and all subsample estimates are similar.  The coefficient on `young' is -0.048 and  the subsample estimates are similar, but the estimates are only marginally significant.  The analysis suggests that  the effect of the dependent population  on inflation is  positive but not statistically well determined,  while there is a negative effect of both `old' and `young' on growth.\footnote{For the inflation regression, the LD estimate for the longest sample is -0.194 with a $t$ statistic of -2.219, and estimates of the shorter subsamples are not statistically significant. The LD longest sample estimate of `old' in the growth regression -0.184 with a $t$ statistic of -4.421, but the subsample estimates are more variable, ranging from -0.051 to -0.451.} A possible explanation for why  the estimate on `old' is better determined  is that the data for 'old' has more variability because its  cross-section variance    fans out.   When there are  two underlying components in the covariate,  using two covariates instead of the sum  can be informative.

% the estimated coefficient  on `old' stablizes around $K=20$ and is well determined,  while the coefficient on the young mirrors that on the dependency ratio. The full sample estimate on 'old is 0.288 and on 'young' is 0.406.  For the growth regression, the full sample estimate on `old' is -0.154 and on `young' is -0.046. Both estimates are reasonably stable with respect to $K$, but the estimates on `old' are  better determined. 

%\begin{figure}[ht!]
%\caption{The Long-Run Relation Between  Demographics and Inflation/Growth}
%\label{fig:infl}
%\centering
%\includegraphics[width=3.20in,height=4.0in]{/do_app02wb_seg.png}
%\includegraphics[width=3.20in,height=4.0in]{/do_app06wb_seg.png}
%\end{figure}

\subsection{Permanent and Transitory Components}
The dependency ratio is a sum of two components that we observe. 
 Suppose now $x_{it}=\mu_{it}+v_{it}$ is a sum of two latent components, where  $\mu_{it}=\mu_{it-1}+u_{it}$  is   non-stationary  and $v_{it}$  is a  mean-reverting process.  As shown earlier, $x_{it}$ can be represented as an IMA(1) process with moving-average parameter $\vartheta$.
%Suppose first that  $w_{it}=\lambda_i f_t$ and $f_t$ is a stationary common component. The cross-section variance is $\var_i(x_{it})=\mathbb E_i[\omega^2_i]t +\var_i(\lambda_i)f_t^2$ which fans out at rate $t$. If there is a large negative common shock at $t$,  the cross-section variance will spike at $t$ but will decline subsequently provided that the non-stationary variations are dominant. 
If  $\mu_{it}$ and $v_{it}$ have the same effect on $y_{it}$, the presence of $v_{it}$  would not change the analysis because $\Delta x_{it}=u_{it}$ already allows for  the possibility that $u_{it}$  can be serially correlated. Nonetheless, as  seen in Table \ref{tbl:csvar},   the cross-section variance  will be small if the variability from $\mu_{it}$ is small. 
%\st{ such as if $\vartheta=-1+\delta/\sqrt{T}$ with $\mathbb E_i[\omega_i^2]=\delta^2/T$ discussed earlier, then $\var_i(\bar X_{iK})$ will no longer fan out, and a $\sqrt{N}$ rate is all that K-averaging can achieve.} 
When the fanning-out effect is limited, regressions in time-compressed I(1) data will not benefit from signal magnification. In that case, omitted serial correlation and fixed effect will no longer be   of second-order importance when the data only permit $\sqrt{N}$ consistent estimation.

A second issue arises if $\mu_{it}$ and $v_{it}$  have distinctive impact of $y$.   Consider
\[y_{it}=\beta_\mu \mu_{it}+\beta_v v_{it}+e_{it}.\]  
As $\cov(x,y)=\beta_\mu \var(\mu)+ \beta_v\var(v)$, 
 the estimand ($\beta_x$) in a regression  of $y$ on  $x$  will be a variance weighted average of $\beta_\mu$ and $\beta_v$. It is as if the regression error   $e_{it}^*= (\beta_v-\beta_\mu) v_{it}+e_{it}$ has a measurement noise. The problem is analogous to the inability to identify the effect of permanent income on consumption from income data. See, for example, \citet{cochrane-94}.  With point-sampling at $T_1$, the weight on $\beta_\mu$ would depend exclusively on the cross-section volatility of $v_{i,t}$ at $t=T_1$. Here, a regression in
  $K$-averaged data  helps because the estimand is
\begin{eqnarray*}
\beta_x=\frac{\cov_i(\bar Y_{iK},\bar X_{iK})}{\var_i(\bar X_{iK})}=\frac{\beta_\mu \var_i(\bar \mu_{iK})+\beta_v \var_i(\bar v_{iK})}{\var_i(\bar \mu_{iK})+\var_i(\bar v_{iK})}.
\end{eqnarray*}
In terms of regression noise, $\bar e_{iK}^*=(\beta_v-\beta_\mu)\bar v_{iK}+\bar e_{iK}=O(K^{-1})$, and in terms of signal,  $\var_i(\bar \mu_{iK})=O(K)$. If $K$ is small, OLS will estimate a weighted average of $\beta_\mu$ and $\beta_v$. But  $\beta_x\rightarrow \beta_\mu$ as $K\rightarrow \infty$ if $\var_i(\bar \mu_{iK})=O(K)$.   A larger $K$ better isolates $\beta_\mu$ by increasing the  permanent-to-transitory ratio $\sigma^2_{\bar \mu_K}/\sigma^2_{\bar v_K}$.

\begin{table}[ht!]
\caption{Simulations Using the Local-Level Model}

\label{tbl:table3}
\begin{center}
  DGP: $y_{it}= \beta_{\mu} \mu_{it} + \beta_v v_{it} + e_{it}$, $(\beta_\mu,\beta_v)=(-1,-2.5)$,
 
Cross-section regression: $\tilde Y_{iK} = \beta_x \tilde X_{iK} + \tilde e_{iK}$,   

$(T,N)=(50,50)$, $(T_0,T_1,T_2)=(10,20,40)$

%Local-Level Model: do-table3-locallevel.m

\begin{adjustbox}{width=\textwidth}

\begin{tabular}{lll|rr|rrr|rrrr|rrr}
 $\sigma_u$ & $\rho $   & $\vartheta$ & \multicolumn{2}{c}{Oracle} & \multicolumn{3}{c}{PS}  & \multicolumn{4}{c}{KA}   & \multicolumn{3}{c}{$LD_K(m,T_0,T_2)$}\\ \hline
& & &      $\hat\beta_\stoc^*$ & $ \hat\beta_\hi^*$ &$T_1$  & $T_2$ & $T$ & $(1,T)$ & $(T_0,T_1)$ & $(T_0,T_2)$ & $(T_1,T_2)$ &   $m=0$ & $m=2$ & $m=5$\\ \hline
\input table3_locallevel_est.tex
\input table3_locallevel_se.tex
\end{tabular}
\end{adjustbox}
\end{center}

\end{table}

To illustrate this point, we conduct a monte-carlo exercise with  10,000 draws of data from the local-level model  for different values of $\sigma_u$ and $\rho$  with  $(\beta_\mu,\beta_v)=(-1.0,-2.5)$.   The estimates of $\beta_x$  and corresponding standard errors  are reported  in Table \ref{tbl:table3}.    A    panel regression using $\mu_{it}$ and $w_{it}$ as covariates is infeasible because these variables  are not observed by the econometrician, but they are observed in simulations. We label these estimates as \textsc{oracle}.  Table \ref{tbl:table3} shows that $\hat\beta_x$ is often close to $\beta_\mu$ when  $\vartheta$ is away from -1, but is contaminated by $\beta_w$  when  $\vartheta$ is close to -1. K-averaging is more capable of producing estimates $\hat\beta_x$ that are closer to $\beta_\mu$ than LD or PS when $\vartheta\ne -1$ .

\begin{figure}[ht!]
\caption{Temperature in Celsius for 48 States in the U.S.}
\label{fig:ustemp}
\hspace*{-1.0in}
\includegraphics[width=8.0in,height=2.50in]{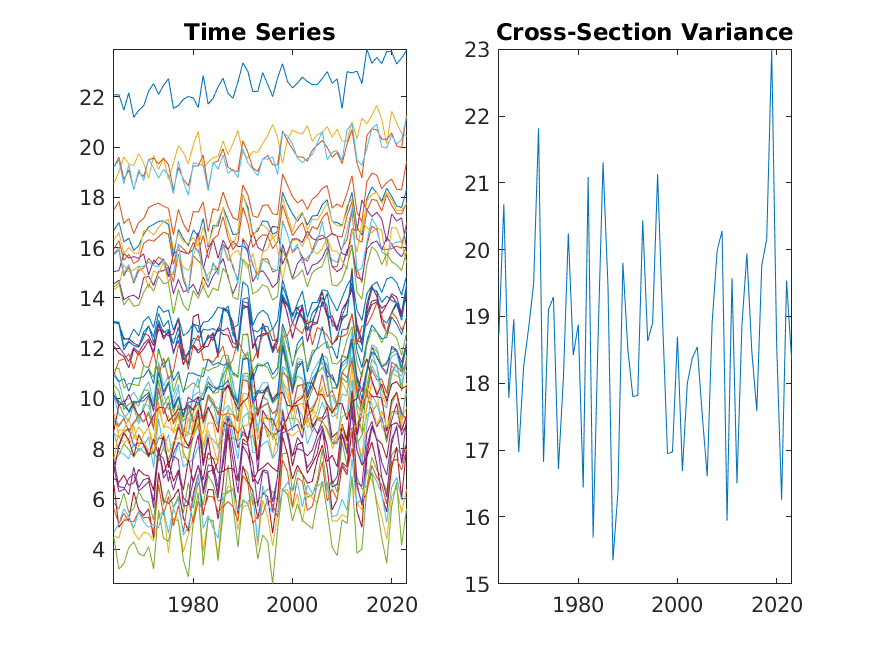}
\end{figure}

Both problems  can be relevant to  regressors such as  temperature if we associate $\mu_{it}$ with the slowly evolving warming trend and $v_{it}$ with regular temperature fluctuations, and itt seems plausible that a warming temperature trend $\mu_{it}$ may affect economic outcome $y_{it}$ differently from regular temperature fluctuations $v_{it}$.
Figure \ref{fig:ustemp} plots the temperature data (in Celsius) for the 48 states in the U.S.\footnote{The latitude data is based on  USGS Geographic Names Information System (GNIS), utilizing the North American Datum of 1983 (NAD 83) aligned with standard WGS 84 latitude-longitude configurations. 
%Country codes via ISO 3166-1 alpha-3; Capital city coordinates via standard World Geodetic System (WGS 84) data listings.
}  There is a slight upward trend in the time series, and yet the cross-section variance hardly fans out. The  limited fanning out  now makes it even more challenging to identify $\beta_\mu$. Unless $\beta_\mu=\beta_v$, a regression of $y$ on $x$ may not return a meaningful parameter estimate, and even time compressed data may not help. 
%\st{Whether this  is due to a small random walk component as in the local-level model, or a small common trend with relatively homogeneous factor loadings as in the case of the factor model, the implication  in terms of the IMA(1) model is that the average long-run variance $\mathbb E_i[\omega^2_i]$ is small.}  

To assess the empirical relevance of the problems, we use output growth as economic outcome; i.e., $y_{it}=100 \Delta \log \text{GSP}_{it}$, where GSP is real state product per capita. As noted in the consumption example, it is useful to  check if the estimates using K-averaged data are stable with respect to $K$.  We first  estimate the linear model   $\bar Y_{iK}=c+ \beta_1 \bar X_{iK}+ $ error and find that  the   estimates   decline as $K$ increases.  Adding latitude of the state capital stabilizes the  estimates, reinforcing what has been suggested in the literature that cross-section regressions can be sensitive to omitted variable bias.

Next, we consider possible non-linear effects   by adding a quadratic term:
\[\bar Y_{iK}= b_0 + \beta_1 \bar X_{iK} + \beta_2 \bar X^2_{iK}  + \text{latitude}_i+\text{ error}.
\]
If $x_{it}=x_{i0}+\sum_{s=1}^t u_{is}$  is a random walk, then $\mathbb E[x_{it}^2]=\mathbb E[x_{i0}^2]+t \sigma^2_i$ but $x_{it}^2$ is not I(2) because it remains non-stationary after differencing.\footnote{In particular,  $\var(\Delta x_{it}^2)=4x_{i,t-1}^2\sigma^2_i+2\sigma_i^4=O(t)$. \citet{ng:08} refers to $x_{it}^2$ as a heteroskedastic random walk; see also  \citet{rico-gonzalo:14}.} Inference would be difficult  for time series regressions. Interestingly, in this data,  $ \var_i(x_{it}^2)$  fans out even  though  $\var_i(x_{it})$ does not. With time compressed data,  standard normal inference is possible  without knowing the exact order of integration.
Note, however, that $\bar X_{iK}^2$ is the average of $x_{it}^2$, not $(\bar X_{ia})^2$ as in  \citet{burke-emerick:16}.  It should also be noted that once the non-linear term is present, the estimates using K-averaged data are similar with or without 'latitude and stable  with respect to $K$.

\begin{figure}[ht!]
\caption{Regressions of Growth on Temperature: 48 States}
\label{fig:climate}

\medskip
%\includegraphics[width=7.250in,height=1.750in]{/do_app010.png}

%\bigskip

\includegraphics[width=6.250in,height=3.50in]{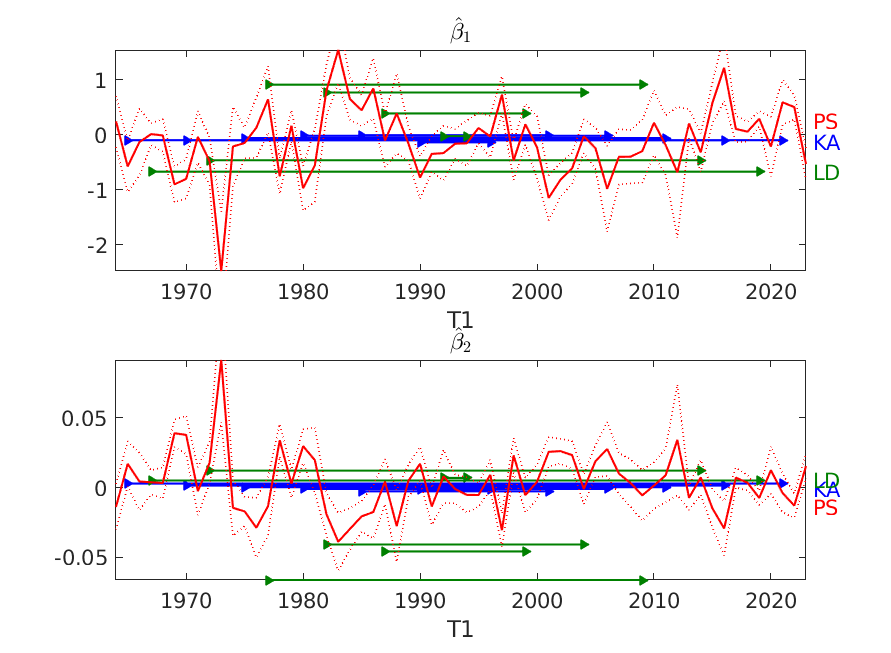}

{\footnotesize See notes to Figure 2.}
\end{figure}

Figure \ref{fig:climate} plots estimates from the non-linear model  for different $(T_0,T_1)=(\kappa,T_1-\kappa)$ pairs with $\kappa$=2:5:30.    The K-averaging estimates (in blue) of $\beta_1$  are negative while the $\beta_2$ estimates are positive. However, the $t$ statistics for full-sample averaging are -1.301 and 1.236, respectively, suggesting that both the linear and  non-linear terms  are not statistically different from zero.  Evidence for a significant long-run relation between growth and temperature  from the sub-sample estimates is even weaker.  In this application, the LD estimates (in green) range from positive to negative, but they are also not well determined. Thus, even with  methodologies that  target the non-stationary variations,  $\beta_\mu$ is not well determined in the data.  

The point sampled estimates help to shed light on the results in this setting. As seen in Figure \ref{fig:climate}, there are  significant fluctuations in the PS estimates (in red) over time, suggesting that  the stationary variations  dominate  the data, consistent with the lack of fanning out in the cross-section variance of temperature. When the non-stationary variation is so weak,  the cross-section  estimate of $\beta_x$ will be close to $\beta_v$ which  a panel regression with time and individual fixed effect targets.   Similarity of the cross-section (long-run) and panel data (short-run) estimates have been interpreted as lack of adaptation  in the literature. But it could also  be that the data does not have enough trending variations to precisely estimate its long-run relation between temperature and growth.
%In principle, it could also be that there is no long-run relation to be uncovered.  This is plausible if  GDP growth is stationary, so that the non-stationary $\mu_{it}$ must either be absent or $\beta_\mu$ is zero for the regression to be balanced. But \citet{gln:26} find a small common trend in this data.  With this data, it seems that we can only meaningfully estimate the shorter term effects of temperature ($v_{it}$) on growth.

\section{Conclusion and Discussion}
 
This paper is built on the result that the cross-section variance of a class of time-compressed I(1) data will fan out, and the magnified signal  can  be exploited to yield (super-)consistent and asymptotically normal estimates. The fastest convergence rate obtains when the compression scheme not only amplifies signal, but also dilutes  the stationary regression noise to render  endogeneity and misspecification bias of second order.  We investigate the long-run relation  between consumption and income, between inflation/growth and demographic variables, and between temperature and growth. While the first two examples find estimates that are stable with respect to $K$, the last example finds that temperature data are dominated by short-term fluctuations, making it difficult even for time compression schemes to isolate the effect of the trending variations.

\clearpage

\renewcommand{\thesection}{\Alph{section}}
\setcounter{section}{0}

\setcounter{section}{0}
\renewcommand{\thesection}{A-\arabic{section}}
\section{Appendix}

In what follows, we will use  $\mathbb{E}_i[\,\cdot\,]$ and $\mathrm{Var}_i(\,\cdot\,)$ to denote
expectation and variance over the cross-sectional distribution of units. To simplify notation, we write $\mathbb E[\cdot |i]$ as $\mathbb E[\cdot]$.

%For the variance, we need
%\begin{eqnarray*}
%E[\int_0^1 W(r)dr]^2&=&\int_0^1 \int_0^1 E[W(s)W(r))dr ds
%= \int_0^1 \int_0^1 \min(s,r)dr ds\\
%&=&\int_0^1 \int_0^s rdrds+\int_0^1\int_s^1 sdrds
%= \int_0^1 \frac{s^2}2ds+\int_0^1 s(1-s)ds\\
%&=&\frac{1}{2} \frac{s^3}{3}\bigg|_0^1+\frac{s^2}{2}\bigg|_0^1-\frac{s^3}{3}\bigg|_0^1=\frac{s^2}{2}\bigg|_0^1-\frac{1}{2}\frac{s^3}{3}\bigg|_0^1 =1/2-1/6=1/3.
%\end{eqnarray*}
%This implies that
%$ \frac{\var(T^{-1/2}\bar X_T)}{\var(T^{-1/2} x_T)}= 1/3.$
%See check-I1var.m

\section*{Proof of Lemma \ref{lem:lemma1}}
If  $x_t$ is stationary, then $\sqrt{T}\bar X\dconv N(0,\omega_x^2)$, where $\omega_x^2=2\pi \sum_{j=-\infty}^\infty \gamma(j)$ is the long-run variance of $x$.
If $\Delta x_t=v_t$, $v_t\sim (0,\sigma^2)$ and is serially correlated with long-run variance $\omega^2_v$, then  $\frac{\bar X}{\sqrt{T}}=T^{-3/2}\sum_{t=1}^T x_t\Rightarrow \omega^2_v\int_0^1 W(r)dr=A$. Since $\mathbb E[\int_0^1 W(r)dr]^2=1/3$,
 $\var(T^{-1/2}\bar X)= \omega_v^2 /3$. 
To understand the finite sample analog of this asymptotic analysis in a general setting, observe that
from  $x_{it}=x_{i,t-1}+u_t$, $x_{it}=x_{i,T_0}+\sum_{s=T_0+1}^t u_{is}$, we have     $\bar X_{i,K}=\frac{1}{K}\sum_{t=T_0+1}^{T_1} x_{it}=x_{i,T_0}+ \frac{1}{K}\sum_{t=T_0+1}^{T_1}\sum_{s=T_0+1}^{t}  u_{is}$. Since $u_{is}$  in the double sum for $t=s$ to $s= T_1$ appears $T_1-s+1$ times,
and $T_1-s+1=K-(s-T_0)+1$, we can let  $j=T_1-s+1$  so that $j=1$ when $s=T_1$ and $j=K$ when $s=T_0+1$   to obtain
\begin{eqnarray*} 
\bar X_{iK}&=&x_{i,T_0}+\sum_{s=T_0+1}^{T_1} \frac{T_1-s+1}{K} u_{is}
= x_{i,T_0}+\sum_{j=1}^{K}\frac{(K-j+1)}{K} u_{i,T_0+j}\\
&=& x_{i,T_0}+\sum_{j=1}^{K} \phi_j u_{i,T_0+j}, \quad\quad \text{where } \quad \phi_j=\frac{K-j+1}{K}.
\end{eqnarray*}
Note  that  $\mu_X=
 \mathbb{E}_i[\bar{X}_{iK}] = \mathbb{E}_i[x_{i,T_0}],
$ because $u_{it}$ have mean zero. Furthermore, $\sum_{j=1}^K j^2=\frac{K(K+1)(2K+1)}{6}$. We thus have $\var(\bar X_{iK})= T_0 \omega^2_i + a_K \omega^2_i$, where $a_K=\frac{(K+1)(2K+1)}{6}$. Therefore, $\mathbb \var_i(\bar X_{iK})= \var_i(x_{i,T_0})+a_K \mathbb E_i[\omega^2_i ]$, where
$\var_i(x_{i,T_0})=\mathbb{E}_i[(x_{i,T_0}-\mu_X)^2]=T_0\mathbb E_i[\omega^2_i]$ .

\section*{Proof of Lemma \ref{lemma:smooth}}
%\paragraph{Proof of Lemma \ref{lem:lemma2}}
Since $x_{it}=x_{it-1}+u_{it}$,  $x_{it}=x_{i,T_0}+\sum_{s=T_0+1}^{t}u_{is}$. The weighted
average $\bar X_i=\sum_t w(t)x_{it}$ with $w_t>0$ and $\sum_t w(t)=1$. More generally,

%As an example, suppose that $T_0=4, T_1=10$ so that $K=6.$  Simple averaging yields
%\begin{eqnarray*}
%\bar X_{iK}&=&\frac{1}{6} x_5+\frac{1}{6} x_6
%+\frac{1}{6} x_7+\frac{1}{6} x_8
%+\frac{1}{6} x_9+\frac{1}{6} x_{10}=\sum_{t=T_0+1}^{T_1} w(t) x_t\\
%&=&x_4+ u_5+ \frac{5}{6}u_6+\frac{4}{6}u_7+\frac{3}{6} u_8+\frac{2}{6} u_9+\frac{1}{6}u_{10}=x_{T_0}+\sum_{s=T_0+1}^{T_1} \phi_{s} u_{s}.
%\end{eqnarray*}

\[
\bar{X}_{iK}
  = \sum_{t=T_0+1}^{T_1} w(t)\bigg(x_{i,T_0}+\sum_{s=T_0+1}^{t}u_{is}\bigg)
  = x_{i,T_0} + \sum_{t=T_0}^{T_1} w(t)\sum_{s=T_0+1}^{t}u_{is}.
\]
Reversing the order of summation to sum over $s$ first, then over
all $t\geq s$, we have
$
\bar{X}_{iK}
  = x_{i,T_0} + \sum_{s=T_0+1}^{T_1}u_{is}\sum_{t=s}^{T_1}w(t).
$
Defining $\phi_s\equiv\sum_{t=s}^{T_1}w(t)$ yields:
\begin{equation}\label{eq:innovation_rep}
\tilde{X}_i = x_{i,T_0} + \sum_{s=T_0+1}^{T_1}\phi_s\,u_{is},
\qquad
\phi_s \equiv \sum_{t=s}^{T_1} w(t),
\end{equation}
where $\phi_s$ is the tail sum of $w$ at $s$, and   the sequence
$\{\phi_s\}$ is the survival function of the weight
distribution $w$. Since $\phi_s \in[0,1]$ for any non-negative $w$ such that $\sum_{t=T_0+1}^{T_1}w(t)=1$, we have $\phi_s^2\le \phi_s$ for each $s$. Consider now \[\sum_{s=T0+1}^{T_1} \phi_s = \sum_{s=T_0+1}^{T_1} \sum_{t=s}^{T_1} w(t) = \sum_{t=T_0+1}^{T_1}w(t) \sum_{s=T_0+1}^t 1 = \sum_{t=T_0+1}^{T_1} w(t) (t-T_0) = K \mathbb E_w\bigg[\frac{t-T_0}{K}\bigg].\] 
 For any non-degenerate $w$,  $\mathbb E_w[(t-T_0)/K]=O(1)$. Thus  $\sum_s \phi_s=O(K)$ and $\sum_{s=T_0+1}^{T_1} \phi_s^2=O(K)$.

In the stationary case,  we have, by definition,
\begin{eqnarray*}
\var(\tilde X_{iK})&=& \sum_{s=T_0+1}^{T_1}\sum_{t=T_0+1}^{T_1} w(t) w(s) \cov(x_{it},x_{is}).
\end{eqnarray*}
Consider  $x_{it}=\rho x_{i,t-1}+u_{it}=\sum_{s=0}^\infty \rho_i^s u_{t-s}$  where $u_{it}\sim (0,\sigma^2_i)$, we have $\bar X_{iK}=\sum_s \psi_s(\rho_i) u_{is}$ with $\psi_s(\rho_i)=\sum_{t=s}^{T_1} w(t) \rho_i^{t-s}$ and
 $\cov(x_{it},x_{is})=\frac{\sigma^2_i}{1-\rho_i^2} \rho_i^{|t-s|}$. Rearranging the double sum by letting $h=s-t$ which runs from $-(K-1)$ to $K-1$:
\begin{eqnarray*}
 \var(\tilde X_{iK})&=&\frac{\sigma^2_i}{1-\rho_i^2} \sum_{h=-(K-1)}^{(K-1)} \sum_{t=-s=h} w(t) w(s) \rho_i^{|t-s|}\\
&=& \frac{\sigma^2_i}{1-\rho_i^2} \sum_{h=(K-1)}^{K-1} A_h \rho_i^{|h|} 
= \frac{\sigma^2_i}{1-\rho_i^2} G(\rho_i)\\
& \equiv &\sigma^2_i \Psi_K(\rho)
\end{eqnarray*}
where $A_h=\sum_{t-s=h} w(t)w(s)$, and
$\Psi_K(\rho)=\frac{1}{1-\rho_i^2} G(\rho_i).$ But $\{A_h\}$ forms a probability distribution over periods separated $h$ periods because  $A_h>0$ and $\sum_h A_h=\sum_h \sum_{t-s=h} w(t) w(s)=\sum_t \sum_s w(t) w(s) =(\sum_t w(t))(\sum_s w(s))=(\sum_t w(t))^2=1$. It follows that, $G(\rho_i)=\mathbb E_A[\rho_i^{|h|}]$. Since $0\le \rho_i^{|h|}\le 1$ for any $|\rho_i|\le 1$, $0\le G(\rho_i)\le 1$ for any $\rho_i\in[0,1]$ and any $w$. 

In the local to unity framework, $\rho_i^{|h|}=(1-c_i/K)^{|h|}\approx 1-\frac{c_i |h|}{K} $. Thus, $G(\rho_i)=\mathbb E_w[ 1-\frac{|h|a_i}{K}]=1-\frac{c_i}{K}\mathbb E_w[|h|]$.  The order of $\Psi_K$ is fundamentally determined by $\rho_i$ and  the choice of $w(t)$, so long as  $w(t)\ge 0$ and $\sum_t w(t)=1$. For simple averaging, the result is immediate since  $\mathbb E_A[|h|]\approx K/3$ and thus $G(\rho_i)\approx 1=O(1)$. We have  $\var(\bar X_{iK})=\frac{1}{K}\frac{\sigma^2_i}{(1-\rho_i)^2}$.   In the local-to-unity case when $\rho_i=1-c_i/K$, $\sigma^2_i=K^2\sigma^2/a_i^2$, implying $\Psi_K=O(K)$. Thus  $\var_i(\bar X_{iK})$ behaves as if $\rho_i=1$, and $Q_{XX}=O(K)$.

\section*{Proof of Proposition \ref{prop:prop1}}
We consider the case of $K$ averaging. 
The OLS estimator satisfies:
\[
\hat\beta - \beta =
  \frac{\frac{1}{N}\sum_i(\bar{X}_{iK}-\bar{\bar{X}})\,\bar{e}_{iK}}
       {\frac{1}{N}\sum_i(\bar{X}_{iK}-\bar{\bar{X}})^2}.
\]
Since $u_{t}\perp e_{is}$ by assumption, the estimator is unbiased at every fixed $K$; i.e., $\cov(\bar X_{iK},\bar e_{iK})=0$.
\paragraph{Part (i): $K$ fixed, }
By LLN, the denominator converges to the fixed constant:
\[
Q_{XX} \equiv \var_i(\bar{X}_{iK})
  = a_K\,\mathbb{E}_i[\sigma_i^2] + \mathbb E_i[var(x_{i,T_0})].
\]
Since $E[u_{is}e_{it}]=0$ for all $t,s$ by assumption,  a CLT applied to   $\frac{1}{\sqrt{N}}\sum_i(\bar{X}_{iK}-\bar{\bar{X}})\bar{e}_{iK}$
 gives a mean zero process  with variance $V_K=\mathbb E_i[(\bar X_{iK}-\mu_X)^2\bar e_{iK}]$.  By law of iterated expectations, $\mathbb E_i[(\bar X_{iK}-\mu_X)^2 \bar e_{iK}^2]=\mathbb E_i\bigg[(\bar X_{iK}-\mu_X)^2\mathbb E_i[\bar e_{iK}^2| \tau_i^2, \bar X_{iK}]\bigg]=\mathbb E_i[(\bar X_{iK}-\mu_X)^2\tau_i^2/K$.  Expanding $(\bar X_{iK}-\mu_X)^2=(\sum_1j \phi_j x_{i,T_0+j}+(x_{i,T_0}-\mu_X))^2$ and using $\mathbb E[u_{is}]=0$,
\begin{eqnarray*}
V_K=\mathbb E_i[(\bar X_{iK}-\mu_X)^2 \bar e_{iK}]&=& \frac{1}{K}\bigg[a_K\mathbb E_i[\sigma^2_i\tau_i^2]+\mathbb E_i[(x_{i,T_0}-\mu_X)^2 \tau_i^2\bigg]
\end{eqnarray*}
 Combining the results for the numerator and denominator gives the variance stated. We keep $\sigma^2_i$ and $\tau_i^2$ inside a joint cross-section expectation since they may be correlated across units. The same applies to $x_{i,T_0}$ and $\tau_i^2$.

\paragraph{ Part (ii): $K\rightarrow \infty$:}  $a_K=(K+1)(2K+1)/(6K)\approx K/3$ and the term relating to  the initial condition becomes  negligible. Now
$Q_{XX}^{-2} V_K =O(K^{-2})$.  Multiplying the estimator by $K$
$\sqrt{N}K (\hat\beta-\beta) \dconv N(0, \mathcal Q_{XX}^{-1} \mathcal V \mathcal Q_{XX}^{-1})$. For this model,   $\mathcal V =\mathbb E_i[\sigma^2_i \tau_i^2]/3$ and $\mathcal Q_{XX}= \mathbb E_i[\sigma^2_i]/3$, giving an asymptotic variance of $3\mathbb E_i[ \sigma_i^2\tau_i^2]/(\mathbb E_i[ \sigma^2_i]) ^2$.

\subsection*{Proof of Proposition \ref{prop:lrm-I1}:}
 The Bewley transformation the ADL(1,1) model yields
\[ y_{it}= \beta x_{it}-\alpha^* \Delta y_{it} -\beta_1^* \Delta x_{it}+e^*_{it},\]
where $\beta=\frac{(\beta_0+\beta_1)}{1-\alpha}$, $\alpha^*=\frac{\alpha}{1-\alpha}$, $\beta_1^*=\frac{\beta_1}{1-\alpha}$, and $e^*_{it}=\frac{e_{it}}{1-\alpha}$. Averaging over $K$ gives
\[ \bar Y_{iK} = \beta \bar X_{iK} -\underbrace{(\alpha^* \Delta y_{iK}+\beta_1^* \Delta x_{iK})}_{\eta_{iK}}+\bar e^*_{iK},\]
where $y_{iK}=\Delta y_{iK}= \frac{(y_{i,T_1}-y_{i,T_0})}{K}=O_p(K^{-1/2})$ and $x_{iK}=\frac{(x_{i,T_1}-x_{i,T_0})}{K}=O_p(K^{-1/2})$ only depend on the endpoint observations.
For fixed $K$,
\[ \plim \hat\beta=\beta -\frac{\cov_i(\bar X_{iK},\eta_{iK})}{\var_i(\bar X_{iK})}\]
  Since $x$ depends on $u$ and $\mathbb E[u_{is}e_{it}]$ for all $s$ and for every $i$, we have $\cov(\bar X_{iK},\bar e_{iK})=0$.  In the I(1) case, $\eta_{iK}=O_p(K^{-1/2})$, which is the same order as $\var_i(\bar e_{iK})$. Furthermore, $\var_i(\bar X_{iK})=O(K)$. By Cauchy Schwarz inequality,
\[ |\cov_i(\bar X_{iK},\eta_{iK})|\le \sqrt{ \var_i(\bar X_{iK})} \sqrt{\var_i(\eta_{iK})}=O(K^{1/2})O(K^{-1/2})=O(1).\]
Thus, bias $\approx O(1)/O(K)=O(K^{-1})$ for fix $K$. 

In the I(0) case, $\Delta x_{iK})=\frac{x_{i,T_1}-x_{i,T_0}} {K}=O_p(K^{-1})$ a4d $\Delta y_{iK}=O_p(K^{-1})$. Thus, $\eta_{iK}=O_p(K^{-1})$. However, $\bar X_{iK}=O_p(K^{-1}$. Thus, bias$\approx \frac{O(K^{-1/2})O(K^{-1})}{O(K^{-1})}=O(K^{-1/2})$. 

In both cases,
\[ \sqrt{N}\bigg(\hat\beta-\beta +\frac{\cov_i(\bar X_{iK},\eta_{iK})}{\var_i(\bar X_{iK})}\bigg)\dconv N(0,Q_{XX}^{-1} W_K Q_{XX}^{-1}),\]
where $W_K=\var_i((\bar X_{iK}-\mu_X)(\bar e_{iK}-\eta_{iK})$. Though $\bar e_{iK}$ does not contribute to bias, it contributes to variance. For large $K$,
\[\frac{ \cov_i(\bar X_{iK},\eta_{iK})}{\var_i(\bar X_{iK})}=\frac{O(1)}{O(K)}=O(K^{-1})\rightarrow 0\]
and with $\mathcal W=\lim_{K\rightarrow\infty} W_K$ and $\mathcal Q_{XX}= \lim_{K\rightarrow\infty} K Q_{XX}$, we have
$ \sqrt{N}K (\hat\beta-\beta)\dconv N(0,\mathcal Q_{XX}^{-1} \mathcal W \mathcal Q_{XX}^{-1}).$

%\section{Proof of Proposition \ref{prop:I0}}
%We now let $x_{it}=\rho_i x_{it-1}+u_{it}$, where $u_{it}\sim (0,\sigma^2_i)$ and $\sigma^2_i=\frac{\sigma^2_i}{(1-\rho_i^2)}$ is the long-run variance of $x_{it}$. It follows from the theory of covariance stationary processes that $\var(\bar X_{iK})=\frac{\sigma^2_i}{K}$, and $\bar X_{iK}=O_p(K^{-1/2})$. 

%For part (i), when the static model is correctly specified, it follows from $\cov_i(\bar X_{iK},\bar e_{iK})=0$ that $\sqrt{N}(\hat\beta-\beta)\dconv N(0,Q_{XX}^{-1} V_K Q_{XX}^{-1})$, where $V_K=\var_i((\bar X_{iK}-\mu_X)\bar e_{iK})$ for all $K$.

%For part (ii), the omitted variables are collected in 
%\[ \eta_{iK}=\frac{\alpha \Delta y_{iK}+\beta_1\Delta x_{iK}}{1-\alpha}=O_p(K^{-1})\]
%since $\Delta x_{iK}=\frac{1}{K}(x_{i,T_1}-x_{i,T_0})=O_p(K^{-1})$, and similarly, $\Delta y_{iK}=O_p(K^{-1})$. Though the bias is shrinking faster than in the I(1) case, $\bar X_{iK}=O_p(K^{-1/2})$, so the omitted variable bias is $OK(K^{-1/2})$, which is larger than in the I(1) case.

\subsection*{Proof of Equation \ref{eq:eqVhomo}}
The meat of the sandwich variance is $V=\var_i(\tilde X_{iK}-\mu_X \tilde e_{iK})$. We want to show that under  homoskedasticity ($\sigma^2_i=\sigma^2, \tau_i^2=\tau^2 \;\forall i$), the cross-sectional variance  is a product of unconditional moments.

Under strict exogeneity, $\mathbb E_i[(\tilde X_{iK}-\mu_X)\tilde e_{iK}]=0$ and $V=\mathbb E_i[ (\tilde X_{iK}-\mu_X)^2 \tilde e_{iK}^2]$. Since $\tilde X_{iK}$ depends on $u_{it}$ and $\tilde e_{iK}$ depends on $e_{it}$, and the shocks are independent across $i$,
\[ \mathbb E_i[(\tilde X_{iK}-\mu_X)^2 \tilde e_{iK}]=\mathbb E_i[(\tilde X_{iK}-\mu_X)^2] \mathbb E_i[\tilde e_{iK}^2]= Q_{XX} \mathbb E_i[\tilde e_{iK}^2].\]
But for a given $i$, $\mathbb E[\tilde e_{iK}]=0$. Thus, $\mathbb E_i[\tilde e_{iK}^2]=\var_i(\tilde e_{iK})=\mathbb E_i[\var(\tilde e_{iK})]+\var_i(\mathbb E[\tilde e_{iK}))$. The second term is zero, and the first term $\mathbb E_i[\var(\tilde e_{iK})]=\mathbb E_i[\tau_i^2 A_0(w)]$. Under homoskdasticty, this is $\tau_2 A_0(w)$. Combining the result, $V=Q_{XX}\tau^2 A_0(w) = \sigma^2 \Phi_K \tau^2 A_0(w)$. The sandwich variance $Q_{XX}^{-1}V Q_{XX}^{-1}$ is thus $\frac{\tau^2}{\sigma^2}\frac{A_0(w)}{\Phi_K}$.

When strict exogeneity fails, $\cov_i(\tilde X_{iK},\tilde e^*_{iK})\ne 0$. Using $\mathbb E_i[A^2B^2]=\mathbb E_i[A^2] \mathbb E_i[B^2]+\cov_i(A^2,B_2)$ with $A=\tilde X_{iK}-\mu_X$ and $B=\tilde e^*_{iK}$, we have, as shown in (\ref{eq:eqV}): 
$ V=\sigma_{e_*}^2 Q_{XX}+\cov_i( (\tilde X_{iK}-\mu_X)^2, \tilde (e^*_{iK})^2).$

\subsection*{Relaxing  Strict Exogeneity}
Under strict exogeneity in Assumption A.i  $, 
\cov_i(\tilde X_{iK},\tilde e_{iK})= \sum_{s}\phi_s \sum_t w(t) \mathbb E_i[u_{it}e_{is}]=0.$ 
We now allow a correlation between $e_{is}$ and $u_{it}$.  Let  $\text{cor}(u_{it},e_{is})=
 \gamma_{i,t-s}$ such that  $\sum_{h=0}^p |\gamma_{i,h}|$ is bounded. In terms of covariance,
$\cov(u_{it},e_{is})=\gamma_{i,t-s} \sigma_i\tau_i.$
From $\tilde X_{iK}=x_{i,T_0}+\sum_s \phi_s u_{is}$ and $\tilde e_{iK}=\sum_t w(t) e_{it}$, 
\begin{eqnarray*}
 \cov(\tilde X_{iK},\tilde e_{iK})&=&\cov\bigg( x_{i,T_0}+ \sum_{s=T_0+1}^{T_1} \phi_s  u_{is}, \; 
\sum_{t=T_0+1}^{T_1}w(t) e_{it}\bigg)\\
&=&
\sigma_i\tau_i\sum_{s=T_0+1 }^{T_1} \sum_{t=T_0+1}^{T_1} \phi_s w(t) \gamma_{i,s-t}\\
&=& \sigma_i \tau_i \sum_{h=-(K-1)}^{K-1} \gamma_{ih} B_h(w), \quad \text{where} \quad B_h(w)=\sum_{s-t=h} \phi_s w(t)
\end{eqnarray*}
where the last equality follows from  grouping by lag $h=s-t$ from $-(K-1)$ to $(K-1)$, for $s,t\in[T_0+1,T_1]$. Note that $B_h(w)$  does not depend on $i$ and let  $B_\infty(w)=\lim_{K\rightarrow\infty} B_h(w)$.
 Since $\sum_h|\gamma_{ih}|<\infty$ and $|B_h(w)|\le \max_s|\phi_s|\sum_t |w(t)|=O(1)$, dominated convergence gives 
\[ \Lambda_K(w,i)\rightarrow \Gamma_i B_\infty(w)\]

For $K$ averaging, $w(t)=1/K$ and $\phi_s=\frac{(T_1-s+1)}{K}$. Thus
\begin{eqnarray*}
B_h(w)= \sum_{s-t=h}\phi_s w(t) = \frac{1}{K} \sum_{\substack{s-t=h\\ T_0+1\le s,t\le T_1}} \frac{T_1-s+1}{K}
=\frac{1}{K}\sum_{j=h+1}^K \frac{(K-j+1)}{K}.
\end{eqnarray*}
When $j=h+1$, the term is $K-h$. When $j=h+2$, it is $K-h-1$, when $j=K$, it is 1. Changing index to let $m=K-j+1$,  we see that  as $j$ increases from $h+1$ to $K$, $m$ decreases from $K-h$ to 1. Thus,
\[ \frac{1}{K}\sum_{j=h+1}^K \frac{(K-j+1)}{K}=\sum_{m=1}^{K-h}m=\frac{1}{K^2}\frac{(K-h)(K-h+1)}{2}\rightarrow 1/2.\]
For point sampling at $T_1$, $w(t)=1_{t=T_1}$ and $\phi_s=\sum_{t>s} w(t)=1$ for all $s\le T_1$. But with any $h$, the only contributing pair is $(s,t)\in[T_1+h,T_1]$ with $\phi_{T_1+h}=1$, and $w(T_1)=1$. Thus, $B_h(w)=B_\infty(w)=1$.  With $\tilde b_K=1/2$ for $K$ averaging, and $\tilde b_K=1$ for sampling at $T_1$, we have $\cov(\tilde X_{iK},\tilde e_{iK})=\sigma_i\tau_i\Lambda_K(w,i)\rightarrow \sigma_i\tau_i B_\infty(w)=O(1)$.

\subsection*{K-averaging when there is a drift}
\begin{itemize}
\item[i.]  Suppose $x_{it}$ is a unit root process  with drift $g_i$; i.e., $\Delta x_{it}=g_i+ u_{it}$ with $\cov(g_i,x_{i,T_0})=0$. Then for $t\in[T_0+1,T_1]$, 
$ x_{it}=x_{i,T_0}+g_i (t-T_0)+\sum_{s=T_0+1}^t u_{is}$
 has both a deterministic and a  stochastic trend. Since $\frac{1}{K} \sum_{t=T_0+1}^{T_1} (t-T_0)=\frac{K+1}{2}$,
$ \bar X_{iK}=x_{i,T_0}+g_i \frac{K+1}{2}+\sum_{s=T_0+1}^{T_1} u_{is}$,
\begin{eqnarray*}
\var_i(\bar X_{iK})&=& \var_i(x_{i,T_0})+ \bar K^2 \var_i(g_i)+ a_K \mathbb E_i[\sigma^2_i]
\end{eqnarray*}
where $\bar K=(K+1)/2$. 
The fanning out rate of $K^2$ is faster when  $g_i\ne 0$ and is heterogeneous.

\item[ii] Suppose now  $\cov_i(x_{i,T_0}, g_i)< 0$.  With convergence,\footnote{We need to evaluate $K^{-2}\sum_s \sum_\ell \cov_i(x_{i,T_0+s},x_{i,T_0+\ell})$. For $[s,\ell]\in [1,K]$, $K^{-2}\sum_s \sum_{\ell} 1=1$, $K^{-2}\sum_s \sum_\ell (s+\ell)=2\bar K$, $K^{-2}\sum_s\sum_\ell s\cdot \ell = \bar K^2$, and $K^{-2} \sum_s \sum_\ell \min(s,\ell)=\bar a_K$.}
\begin{eqnarray*}
\label{eq:Kavg-convgence}
\var_i(\bar X_{iK})&=&\underbrace{ \var_i(x_{i,T_0}) +  2\bar K \cov_i(x_{i,T_0},g_i)+ \bar K^2 \var_i(g_i)+ \bar K \mathbb E_i[\sigma^2_i] }_{\var_i(\bar x_{i,T_0+\bar K})}+(\bar a_K -\bar K) \mathbb E_i[\sigma^2_i].
\end{eqnarray*}
The first three terms represent the cross-section variance at the midpoint of $\bar K$ which is strictly positive,  while the last term  comes from noise accumulation since $\bar K$. Like point sampling, the cross-section variance of the K-averaged data also shows a U-shape and the minimum is generally at a point later than under point sampling.

\item[iii] If $ x_{it}=\lambda_i f_t + w_{it}$, $f_t=g_f +f_{t-1}+u_{it}$, $w_{it}$ is stationary. Then after cross-section demeaning,
\[ \var_i(\bar X_{iK})= \var_i(\lambda_i) \bar f_K^2 +\var_i(\bar w_{iK}).\]
Now $\bar w_{iK}=O(K^{-1/2})$. With  $\bar f_K= f_0+g_f K/2+O(K^{1/2}$, we have 
\[ \bar f_K^2= (f_0+  g_f  K/2)^2 + O_p(K).\] If $g_f<0$, $\var_i(\bar X_{iK})$ can be $U$-shaped. The deterministic part reaches zero at $K^*=-2f_0/g_f=2 t^*$, where $t^*$ is the turning point under point sampling. At $K^*$, $\var_i(\bar X_{iK})=O_p(K^*)$. As with point sampling, a U-shaped cross-section requires $f_0\ne 0$. It
also  does not mean that $\var_i(\bar X_{iK})$ tends to zero, but that recovery can be expected in time. 
\end{itemize}

\clearpage
%\bibliography{../../../climate/write/climate,metrics2,factor,macro,metrics,consum}

\input{gn26.bbl}
\end{document}

%% file: table1_locallevel.tex
3.000 &-0.091 &90.913 &271.069 &451.693 &34.815 &94.577 &154.725 &65.012 &184.669 &305.055 &227.125 &211.202 &180.604 &\\
1.000 &-0.381 &10.981 &31.010 &51.078 & 3.958 &10.537 &17.208 & 7.399 &20.573 &33.930 &27.003 &23.826 &20.185 &\\
0.500 &-0.608 & 3.492 & 8.503 &13.521 & 1.065 & 2.659 & 4.317 & 1.999 & 5.191 & 8.512 & 8.244 & 6.257 & 5.146 &\\
0.200 &-0.817 & 1.397 & 2.201 & 3.005 & 0.255 & 0.453 & 0.707 & 0.487 & 0.886 & 1.396 & 2.994 & 1.337 & 0.935 &\\
0.100 &-0.903 & 1.099 & 1.300 & 1.503 & 0.139 & 0.138 & 0.192 & 0.272 & 0.271 & 0.379 & 2.244 & 0.634 & 0.334 &\\
0.050 &-0.949 & 1.025 & 1.075 & 1.128 & 0.110 & 0.059 & 0.063 & 0.218 & 0.117 & 0.125 & 2.057 & 0.458 & 0.183 &\\
\hline 

%% file: table1_factormodel.tex
4.000 &-0.056 &165.591 &490.136 &821.657 &63.022 &173.429 &282.459 &117.840 &336.314 &552.694 &404.454 &377.922 &322.043 &\\
2.000 &-0.171 &42.142 &123.273 &206.158 &15.831 &43.382 &70.629 &29.611 &84.127 &138.200 &102.621 &94.783 &80.613 &\\
0.800 &-0.457 & 7.580 &20.558 &33.824 & 2.617 & 6.969 &11.317 & 4.906 &13.515 &22.144 &18.101 &15.502 &13.011 &\\
0.600 &-0.552 & 4.700 &12.000 &19.463 & 1.516 & 3.935 & 6.374 & 2.848 & 7.631 &12.473 &11.057 & 8.895 & 7.377 &\\
0.400 &-0.670 & 2.644 & 5.887 & 9.206 & 0.730 & 1.767 & 2.844 & 1.377 & 3.428 & 5.565 & 6.024 & 4.175 & 3.353 &\\
0.200 &-0.817 & 1.410 & 2.221 & 3.053 & 0.258 & 0.467 & 0.726 & 0.494 & 0.907 & 1.421 & 3.004 & 1.343 & 0.938 &\\
\hline 

%% file: table_thm1_est.tex
-1.00 & 0.00 &-1.001 &-0.998 &-0.998 &-1.000 &-1.000 &-1.000 &-1.000 &-0.999 &-0.999 &-1.000\\
-0.99 & 0.00 &-0.989 &-0.990 &-0.988 &-0.990 &-0.989 &-0.990 &-0.990 &-0.990 &-0.990 &-0.990\\
-0.97 & 0.00 &-0.976 &-0.978 &-0.974 &-0.975 &-0.976 &-0.975 &-0.975 &-0.979 &-0.977 &-0.975\\
-0.95 & 0.00 &-0.948 &-0.951 &-0.949 &-0.949 &-0.949 &-0.949 &-0.950 &-0.954 &-0.950 &-0.951\\
-1.00 &-0.50 &-0.996 &-1.001 &-1.005 &-1.001 &-1.001 &-1.002 &-1.001 &-1.003 &-1.000 &-0.998\\
-0.99 &-0.50 &-0.990 &-0.990 &-0.992 &-0.989 &-0.989 &-0.989 &-0.988 &-0.992 &-0.990 &-0.991\\
-0.97 &-0.50 &-0.974 &-0.976 &-0.976 &-0.976 &-0.977 &-0.976 &-0.976 &-0.966 &-0.975 &-0.972\\
-0.95 &-0.50 &-0.954 &-0.946 &-0.948 &-0.950 &-0.950 &-0.949 &-0.949 &-0.951 &-0.947 &-0.951\\
\hline 

%% file: table_thm1_tstat.tex
-1.00 & 0.00 & 0.076 & 0.076 & 0.077 & 0.072 & 0.077 & 0.076 & 0.076 & 0.077 & 0.079 & 0.072\\
-0.99 & 0.00 & 0.077 & 0.074 & 0.076 & 0.094 & 0.076 & 0.090 & 0.084 & 0.075 & 0.081 & 0.074\\
-0.97 & 0.00 & 0.077 & 0.077 & 0.079 & 0.168 & 0.092 & 0.142 & 0.137 & 0.077 & 0.083 & 0.087\\
-0.95 & 0.00 & 0.086 & 0.094 & 0.100 & 0.451 & 0.144 & 0.357 & 0.314 & 0.081 & 0.129 & 0.126\\
-1.00 &-0.50 & 0.076 & 0.072 & 0.076 & 0.072 & 0.077 & 0.074 & 0.080 & 0.071 & 0.075 & 0.079\\
-0.99 &-0.50 & 0.078 & 0.076 & 0.076 & 0.081 & 0.077 & 0.081 & 0.080 & 0.077 & 0.076 & 0.078\\
-0.97 &-0.50 & 0.078 & 0.073 & 0.075 & 0.099 & 0.080 & 0.087 & 0.093 & 0.075 & 0.081 & 0.082\\
-0.95 &-0.50 & 0.082 & 0.086 & 0.081 & 0.183 & 0.096 & 0.152 & 0.142 & 0.082 & 0.090 & 0.089\\
\hline 

%% file: mc_app05_bkn_sample.tex
  mean &  0.962 &  0.946 &  0.961 &  0.949 &  0.949 &  0.951 &  0.930 \\
    se &  0.003 &  0.003 &  0.001 &  0.038 &  0.024 &  0.031 &  0.049 \\
    sd &  0.008 &  0.016 &  0.008 &  0.042 &  0.025 &  0.034 &  0.052 \\
  bias &  0.012 & -0.003 &  0.012 & -0.000 & -0.001 &  0.001 & -0.019 \\
  rmse &  0.008 &  0.016 &  0.008 &  0.041 &  0.025 &  0.034 &  0.052 \\

%% file: table3_locallevel_est.tex
&&&& \multicolumn{10}{c}{Estimates of $\beta_x$} \\ \hline
 4.00 & 0.00 &-0.06 &-1.000 &-2.506 &-1.004 &-1.002 &-1.002 &-1.000 &-1.001 &-1.000 &-1.000 &-1.006 &-1.001 &-1.000\\
 1.00 & 0.00 &-0.38 &-1.000 &-2.502 &-1.071 &-1.037 &-1.030 &-1.002 &-1.012 &-1.003 &-1.003 &-1.093 &-1.012 &-1.006\\
 0.50 & 0.50 &-0.79 &-1.001 &-2.502 &-1.315 &-1.177 &-1.145 &-1.027 &-1.139 &-1.038 &-1.041 &-1.393 &-1.152 &-1.092\\
 0.01 & 0.50 &-1.00 &-1.104 &-2.501 &-2.498 &-2.495 &-2.491 &-2.491 &-2.498 &-2.494 &-2.494 &-2.491 &-2.501 &-2.501\\
\hline 

%% file: table3_locallevel_se.tex
&&&& \multicolumn{10}{c}{S.E.} \\ \hline
 4.00 & 0.00 &-0.06 & 0.007 & 0.250 & 0.027 & 0.019 & 0.017 & 0.004 & 0.010 & 0.005 & 0.005 & 0.031 & 0.011 & 0.008\\
 1.00 & 0.00 &-0.38 & 0.028 & 0.086 & 0.106 & 0.076 & 0.068 & 0.017 & 0.041 & 0.020 & 0.021 & 0.121 & 0.044 & 0.033\\
 0.50 & 0.50 &-0.79 & 0.079 & 0.070 & 0.195 & 0.149 & 0.134 & 0.041 & 0.094 & 0.048 & 0.051 & 0.217 & 0.100 & 0.077\\
 0.01 & 0.50 &-1.00 & 7.909 & 0.055 & 0.377 & 0.381 & 0.380 & 0.222 & 0.234 & 0.222 & 0.227 & 0.380 & 0.236 & 0.226\\
\hline 